# TWICEBEE: A Two-stage Intra-patient Curve-free Bayesian Decision-Theoretic Dose Escalation Design

Dehua Bi, PhD[1,2]

Katherine Ryan, DO[3]

Sabine Heitzeneder, MD[4]

Zina Good, PhD[4]

John S. Tamaresis, PhD[1]

Robert Lowsky, MD[5]

Michelle Monje, MD, PhD[4]

Crystal Mackall, MD[3,4,5,6,7]

Ying Lu, PhD[1]

1. Department of Biomedical Data Science and Center for Innovative Study Design, School of Medicine, Stanford University, Stanford, CA 94305, USA
2. Center for Cancer Cell Therapy, Stanford Cancer Institute, Stanford University, Stanford, CA 94305, USA
3. Division of Hematology & Oncology, Stem Cell Transplantation and Regenerative Medicine, Dept. of Pediatrics, Stanford University School of Medicine, Stanford University, Stanford, CA 94305, USA
4. Division of Immunology and Rheumatology, Center for Biomedical Informatics Research, Department of Medicine, School of Medicine, Stanford University, Stanford, CA 94305, USA
5. Division of Bone Marrow Transplant & Cell Therapy, Dept. of Medicine, Stanford University School of Medicine, Stanford University, Stanford, CA 94305, USA
6. Parker Institute for Cancer Immunotherapy, San Francisco, CA 94129
7. Weill Cancer Hub West, Stanford, CA 94305

Corresponding author: Ying Lu, PhD, Professor of Biomedical Data Science, Stanford University School of Medicine, Stanford, CA94305-5464, USA. E-mail: ylu1@stanford.edu; Tel: 650-736-8300.

**Abstract**

We propose a novel Phase I intra-patient dose-escalation design tailored for multi-cycle immunotherapy settings, in which toxicity at a fixed dose level is clinically expected to decrease over successive treatment cycles. This design was motivated by a phase I trial of CAR T-cell therapy, an emerging cellular immunotherapy with established applications in cancer and growing investigation in autoimmune disease. The design is intended for settings in which nonincreasing cycle-specific toxicity assumption is clinically justified. Specifically, we build on the extrapolation property of the modified curve-free Bayesian decision-theoretic (c-CFBD) design for two-agent trials (Xu, et al. 2025), treating treatment cycle as a second dimension. By redefining the partial order, the c-CFBD framework can accommodate the reduction in toxicity across cycles. The proposed design adopts a two-stage structure: an initial accelerated titration stage to rapidly explore dose levels, followed by a c-CFBD stage to improve safety and estimate the cycle-specific maximum tolerated dose sequence. Simulation studies across a range of scenarios demonstrate favorable operating characteristics.

# 1. Introduction

In oncology, Phase I clinical trials primarily evaluate the safety of new treatments. Patients are enrolled one after another. Dose levels are assigned to individual patients or cohorts based on a prespecified algorithm. The resulting data serve following main purposes: (1) guiding dose assignments for future patients, (2) estimating the maximum tolerated dose (MTD) or maximum administered dose (MAD) once the trial meets its stopping criteria, (3) recommended phase 2 dose (RP2D) after consideration of MTD, pharmacokinetics and pharmacodynamic biomarkers, and preliminary efficacy signals.

For decades, the 3+3 design was the predominant approach for assigning dose levels in Phase I trials. This algorithmic method was widely used. However, over the past decade, the field has moved away from the 3+3 design due to its inferior performance, including a tendency to allocate patients to subtherapeutic doses (Ji and Wang 2013). Modern alternatives include model-based designs such as the continual reassessment method (CRM) (O'Quigley, Pepe and Fisher 1990) and model-assisted designs such as mTPI (Ji and Wang 2013), mTPI-2/Keyboard (Guo, et al. 2017), and BOIN (Liu and Yuan 2015). Algorithmic designs such as i3+3 (Liu, Wang and Ji 2020) have also been developed. Other approaches include the curve-free Bayesian decision-theoretic (CFBD) design (Fan, Lu and Wang 2012) (Lee, Fan and Lu 2017) and the c-CFBD design (Xu, et al. 2025).

Each modern design has distinct advantages. However, all share a common assumption: dose assignment occurs only once at a patient’s initial enrollment, and only the outcome from that first treatment cycle informs subsequent dose decisions. This assumption does not hold for multi-cycle regimens, where dose decisions may need to be made repeatedly for the same patient in different treatment cycles.

Not all multi-cycle regimens share the same toxicity pattern across treatment cycles. In this paper, we focus on the setting in which, at a fixed dose, toxicity is expected to be highest in the initial cycle and to decrease over later cycles. Our motivating example comes from intracranial CAR T-cell therapy, but the same structure may arise more broadly in multi-cycle immune- and cell-therapy settings. Thus, the proposed design is not intended as a universal intra-patient dose-escalation design for all oncology regimens. Rather, it is developed for an important and increasingly relevant class of trials in which a nonincreasing cycle-specific toxicity pattern is clinically justified.

### 1.1. Motivating Trial

Our motivating example is a Phase I trial of intracranial chimeric antigen receptor (CAR) T-cell therapy in patients with H3K27M-mutated diffuse midline glioma (DMG). Specifically,
a Phase I, open-label, single-site trial (IND19801; NCT04196413) is evaluating autologous GD2 CAR-T cells in this patient population (Monje, et al., 2025). The trial aims to assess the feasibility of manufacturing GD2 CAR T-cells, evaluate safety, and determine the RP2D/MTD, route, and schedule of GD2 CAR T-cells. Because each patient may receive multiple treatment cycles, the trial also raises the question of how doses should be selected not only for the initial administration, but also for subsequent administrations within the same patient.

Arm A of the trial initially used the 3+3 design to determine the MTD for the first cycle. CAR T-cells were administered by intravenous (IV) injection following lymphodepleting chemotherapy.

Patients who demonstrated clinical or imaging benefit were eligible to receive subsequent intracerebroventricular (ICV) infusions. In the published Arm A experience, dose-limiting toxicities occurred at the higher IV dose level, whereas no dose-limiting toxicities were reported among the subsequent ICV infusions (Monje, et al. 2025). These results suggest that repeated intracranial administration is clinically feasible in selected patients, while also highlighting the importance of careful toxicity monitoring over multiple administrations.

The trial was later expanded to include ICV-focused treatment arms. These arms evaluate GD2 CAR T-cells administered intracerebroventricularly, with different conditioning strategies. The ongoing study is also described as using a rolling-6 escalation design. Therefore, this motivating trial represents a setting in which dose, route, schedule, conditioning strategy, and repeated administration are all relevant to early-phase safety evaluation.

Preliminary results from Arm A demonstrated that toxicity was greatest following the first dose and diminished with subsequent doses. This pattern was observed for tumor inflammation-associated neurotoxicity, whose grade typically decreased over later infusions (Monje, et al. 2025, Majzner, et al. 2024). Such a toxicity profile is clinically plausible in multi-cycle immunotherapy settings, where early treatment may occur in the presence of higher tumor burden and stronger inflammatory response. It also suggests that the toxicity risk at a fixed dose may not remain constant across cycles. Instead, for some therapies, the same dose may be less toxic in later cycles than in the initial cycle.

This observation motivates a modified intra-patient dosing strategy. In this setting, patients may benefit when doses can vary across cycles, for example by escalating to a higher dose level in later cycles. At the same time, safety remains the primary concern in Phase I trials. An effective design should minimize the number of patients treated at overly toxic dose-cycle combinations that may lead to unacceptable risk of DLT. For example, if two out of three patients experience DLTs at dose level 3 in cycle 3, it may be inappropriate to assign another patient to that same dose-cycle combination.

Taken together, the preliminary trial experience suggests a setting in which, for a fixed dose, toxicity is highest in the initial cycles and decreases over subsequent cycles. This motivates an intra-patient dose-escalation design that uses information from all treated patients to make cycle-specific dose decisions. Because patients receive treatment over multiple cycles and toxicity may change over time, the design should target not a single MTD, but a sequence of cycle-specific MTDs. Such an approach is better suited to identifying a safe and potentially efficacious dosing strategy across the full treatment course.

The relevance of this setting extends beyond a single intracranial CAR-T trial, as repeated dosing strategies are being explored across multiple CAR-T platforms — including locoregional and intracerebroventricular (ICV) delivery — where toxicity monitoring and dose decisions must be made over multiple administrations within the same patient.(Brown, et al. 2024, Vitanza, et al. 2025). More broadly, related early-cycle toxicity patterns have also been reported in other T-cell–redirecting immunotherapies, such as bispecific T-cell engagers and bispecific antibodies, where immune-mediated toxicities are often most pronounced in the initial administrations and may attenuate in later cycles under step-up or cycle-dependent dosing strategies (Kim, et al. 2025, Ahn, et al. 2023). The proposed design is therefore not intended as a universal intra-patient dose-escalation framework for all oncology regimens, but rather is developed for an important and increasingly relevant class of multi-cycle immunotherapy trials in which a nonincreasing cycle-specific toxicity pattern is clinically anticipated and justified.

## 1.2. Intra-patient Dose Escalation Design

Few designs in the literature address intra-patient dosing strategies for novel therapies like CAR T-cell therapy. One of the earliest examples is the accelerated titration (AT) design (Simon, et al. 1997). The AT design allows within-patient dose modification. A patient may escalate to a higher dose, stay at the same dose, or de-escalate to a lower dose based on the toxicity grade observed in the previous cycle. Specifically, the dose changes if toxicity is grade $\leq 1$, grade 2, or $>$ grade 2, respectively. However, the AT design does not use toxicity data from other patients when determining doses for subsequent cycles. This limitation reduces its ability to make fully informed decisions.

Lyn et al. (Lyu, Curran and Ji 2018) proposed the BaSyc design for a Phase I trial in a rare leukemia population. This trial combined the Janus kinase inhibitor ruxolitinib with a three-cycle FDA-approved chemotherapy backbone that included cycle-specific drugs. In BaSyc, Bayesian models determine whether a patient should proceed to the next cycle. If a patient experiences a DLT in the previous cycle, the design calls for de-escalation. Otherwise, the mTPI-2 design (Guo, et al. 2017) selects the most appropriate dose based on accumulated patient data. The BaSyc design is safe and borrows information across patients. However, it is limited to trials with at most three cycles per patient. In CAR T-cell therapy, patients may receive multiple cycles of treatment. This could create an unmanageably large number of dose-cycle combinations and potentially make BaSyc impractical for this setting.

The DIETE design, proposed by Xu et al. (Xu, Zhuo and Rasmussen 2021), targets T-cell engager oncology drugs. It assumes that, for a given dose, toxicity differs before and after the patient adapts to treatment. The goal is to identify the MTD before adaptation (in initial cycles) and the MTD after adaptation (in later cycles). The BOIN design guides dosing in the initial cycles, while subsequent cycles use a beta-binomial model combined with algorithmic rules. Although both T-cell engagers and CAR T-cell therapies are immunotherapies, they differ in mechanism and manufacturing. This can lead to distinct toxicity profiles.

More recently, Zhou et al. (Zhou, et al. 2023) proposed the AIDE design, and Guo and Liu (Guo and Liu 2025) introduced the IP-CRM design. Both are general intra-patient frameworks. These designs improve efficiency by using repeated observations from individual patients to guide dosing and reduce the total sample size. However, neither design accounts for situation of CAR T-cell therapy, where toxicity often decreases over cycles, with the highest toxicity typically observed in the initial cycle. In addition, both designs aim to identify a single MTD, rather than a sequence of MTDs for each treatment cycle.

### 1.3. Proposed Design: TWICEBEE

To address the need for an intra-patient dose-escalation strategy in modern CAR T-cell therapy, we propose the TWICEBEE design. TWICEBEE stands for TWo-stage Intra-patient Curve-frEe Bayesian dEcision-thEoretic dose escalation. The design consists of two sequential stages:

- **Stage 1:** Accelerated titration with a fixed dosing scheme.
- **Stage 2:** Intra-patient dose escalation guided by the modified c-CFBD design.

The trial begins in Stage 1. In this stage, all patients receive treatment according to the same predetermined dosing sequence. The design transitions to Stage 2 when either (i) the first DLT is observed in any patient during any treatment cycle, or (ii) the number of enrolled patients reaches a prespecified maximum initial sample size.

In Stage 2, dosing decisions are determined using a modified version of the two-agent c-CFBD design (Lee, Fan and Lu 2017). This model considers two relationships: a monotone dose-toxicity curve within each cycle, and an inverse-monotone cycle-toxicity curve at each dose level. This framework allows for cycle-specific dosing for both new patients and those already in treatment. It also accounts for the typical decrease in toxicity probability across successive CAR-T treatment cycles.

The remainder of the paper is organized as follows. Section 2 describes the proposed two-stage design in detail. Section 3 illustrates its application using a trial motivated by our case study. Section 4 presents simulation-based operating characteristics and compares TWICEBEE with the DIETE design, the IP-CRM design, and a variant of TWICEBEE in which the modified c-CFBD component is replaced by the i3+3 design. Section 5 concludes the paper with a discussion.

# 2. Method

We first briefly review the c-CFBD design, which provides the main model-based component of the proposed framework. We then describe the proposed modification for the CAR T-cell therapy setting. Finally, we present the full two-stage TWICEBEE design.

## 2.1. Brief Review of the c-CFBD Design for Two-Agent Trials

The calibrated curve-free Bayesian decision-theoretic (c-CFBD) design (Xu, et al. 2025) is a modification of the original curve-free Bayesian decision-theoretic (CFBD) design proposed by Fan et al. (Fan, Lu and Wang 2012) and Lee et al. (Lee, Fan and Lu 2017). In the two-agent setting, c-CFBD uses an isotonic working model with conjugate Beta priors and a partial ordering structure on dose combinations. Below, we summarize only the components needed for the proposed design.

Let $I$ and $J$ denote the numbers of dose levels for agents 1 and 2, respectively. Let $p_{ij}$ be the probability of a DLT at dose combination $(i, j)$, where $i \in \{1, \dots, I\}$ and $j \in \{1, \dots, J\}$. Let $y_{ij}$ be the number of observed DLTs, and $n_{ij}$ the number of patients treated at dose-level combination $(i, j)$. The c-CFBD design assumes

$$y_{ij} | n_{ij}, p_{ij} \sim Binom(n_{ij}, p_{ij}), \qquad p_{ij} \sim Beta(\alpha_{ij}, \beta_{ij}),$$

where $\alpha_{ij}$ and $\beta_{ij}$ are prior hyperparameters that are updated as the trial progresses.

Because each agent has multiple dose levels, the toxicity ordering among dose combinations is only partially known. We write $(i, j) \prec (i', j')$ if the toxicity at $(i, j)$ is lower than that at $(i', j')$, and $(i,j) \succ (i',j')$ for the reverse. In the standard two-agent setting, $(i,j) \prec (i',j')$ if $i \le i'$ and $j \le j'$. Dose combinations that satisfy neither relationship are regarded as incomparable.

After each cohort $k$ is observed, the parameters $a_{ij}^{(k)}$ and $b_{ij}^{(k)}$ are updated from their previous values $a_{ij}^{(k-1)}$ and $b_{ij}^{(k-1)}$ as:

$$\left(a_{ij}^{(k)}, b_{ij}^{(k)}\right) = \begin{cases} \left(a_{ij}^{(k-1)}, b_{ij}^{(k-1)} + n_{i'j'} - y_{i'j'}\right) & (i,j) \prec (i',j'), \\ \left(a_{ij}^{(k-1)} + y_{i'j'}, b_{ij}^{(k-1)} + n_{i'j'} - y_{i'j'}\right) & (i,j) = (i',j'), \\ \left(a_{ij}^{(k-1)} + y_{i'j'}, b_{ij}^{(k-1)}\right) & (i,j) \succ (i',j'), \\ \left(a_{ij}^{(k-1)}, b_{ij}^{(k-1)}\right) & \text{otherwise,} \end{cases}$$

where $(i',j')$ is the dose combination assigned to cohort $k-1$. Thus, dose combinations considered more toxic than the current one receive pseudo-DLT information, whereas those considered less toxic receive pseudo-non-DLT information.

The c-CFBD design calibrates these pseudo counts to equalize the effective sample size (ESS) across all dose combinations. The calibrated hyperparameters are given by

$$\alpha_{ij}^{(k)} = \frac{a_{ij}^{(k)}}{l_{ij}^{(k)}}, \qquad \beta_{ij}^{(k)} = \frac{b_{ij}^{(k)}}{l_{ij}^{(k)}}$$

where

$$l_{ij}^{(k)} = \frac{a_{ij}^{(k)} + b_{ij}^{(k)}}{S^{(k)}}, \qquad S^{(k)} = \frac{\sum_{i=1}^{I} \sum_{j=1}^{J} \left(a_{ij}^{(k)} + b_{ij}^{(k)}\right)}{I \times J}.$$

Here, $S^{(k)}$ is the average ESS. This calibration reduces the over-shooting and under-shooting problem of the original CFBD design. Details can be found in Xu et al. (Xu, et al. 2025).

Dose selection is based on a one-step-look-ahead (OSLA) rule with utility function

$$u_k\left(p_{ij}^{(k)}, p_t\right) = \begin{cases} -\eta^-\left(p_t - p_{ij}^{(k)}\right) & \text{if } p_{ij}^{(k)} \le p_t \\ -\eta^+\left(p_{ij}^{(k)} - p_t\right) & \text{if } p_{ij}^{(k)} > p_t \end{cases},$$

where $p_t$ is the target toxicity rate, and $\eta^-$ and $\eta^+$ are positive constants controlling the penalties for under-dosing and over-dosing, respectively. The next dose combination is chosen as

$$\arg\max_{(i,j)} E\left[u_k\left(p_{ij}^{(k)}, p_t\right) \mid Data_k\right],$$

where $Data_k$ includes all data accumulated up to cohort $k$.

These are the main components of c-CFBD used in the proposed design. Additional technical details, including the original stopping rules and the choice of initial values satisfying monotonicity, are given in Xu et al. (2025).

## 2.2. Proposed Modification of the c-CFBD Design for CAR T-cell Therapy

We now adapt the c-CFBD design to the multi-cycle CAR T-cell therapy setting. The key modification is that dose level is treated as the first dimension and treatment cycle as the second dimension. However, unlike the standard two-agent setting, the toxicity relationship along the cycle dimension is reversed. Specifically, we consider settings in which, for a fixed dose level, toxicity is expected to be highest in the early treatment cycles and to decrease over later cycles. The proposed modification is developed for this setting and should be used only when there is adequate clinical justification for this assumption. In the motivating intracranial CAR-T setting, preliminary experience suggested that toxicity was greatest after the initial administration and tended to decrease in later cycles. A similar pattern may also arise in other multi-cycle immunotherapy settings.

Let $j = 1, \dots, J$ denote dose levels and $c = 1, \dots, C$ denote treatment cycles. For a fixed cycle $c$, toxicity increases with dose:

$$p_{jc} \leq p_{j'c} \text{ if } j \leq j'.$$

For a fixed dose level $j$, toxicity is assumed to be the highest in the first cycle and to decreases in later cycles:

$$p_{jc} \geq p_{jc'} \text{ if } c \leq c'.$$

Thus, when cycles are treated as the second dimension, the toxicity order is reversed relative to the standard c-CFBD design.

**C Partial Order**: To accommodate this difference, we define the "C partial order." For dose-cycle combinations $(j, c)$ and $(j', c')$:

- $(j, c) \prec (j', c')$ meaning $(j, c)$ is less toxic than $(j', c')$, if $j \leq j'$ and $c' \leq c$;
- $(j, c) \succ (j', c')$ meaning $(j, c)$ is more toxic than $(j', c')$, if $j \geq j'$ and $c' \geq c$.

This ordering captures the monotonic increase in toxicity with dose level and the monotonic decrease in toxicity with cycle number. It also allows partial comparability across dose-cycle combinations. Figure 1 provides an illustration.

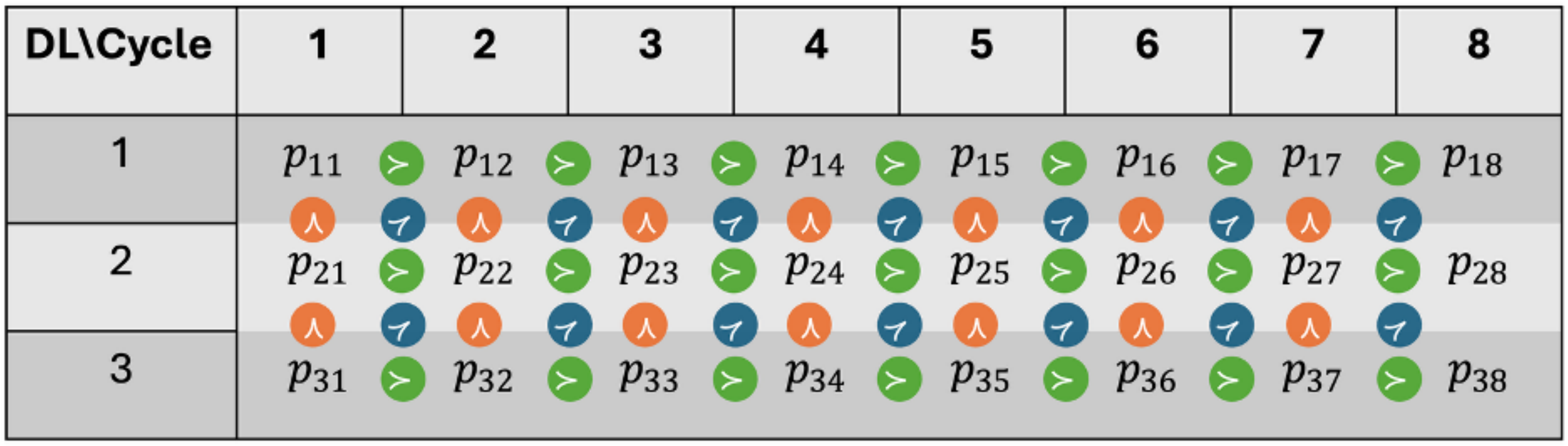

**Figure 1:** Illustration of the C partial order with $J = 3$ dose levels and $C = 8$ cycles. Green arrows indicate ordering across cycles within a fixed dose level, orange arrows indicate ordering across dose levels within a fixed cycle, and blue arrows indicate ordering across both dose and cycle dimensions. The figure illustrates the ordering structure only.

To illustrate how information propagates under the C partial order, consider the setting in Figure 1 and suppose a cohort of three patients is treated at dose-cycle combination (2,4), with $n_{24} = 3$ and $y_{24} = 1$. Under the C partial order, the combinations more toxic than (2,4), such as (3,2), receive pseudo-DLT information. Thus,

$$a_{32}^{(k)} = a_{32}^{(k-1)} + 1, \;\; b_{32}^{(k)} = b_{32}^{(k-1)}.$$

Dose-cycle combinations less toxic than (2,4), such as (1,6), receive pseudo-non-DLT information, thus,

$$a_{16}^{(k)} = a_{16}^{(k-1)}, \;\; b_{16}^{(k)} = b_{16}^{(k-1)} + 2.$$

For the treated combination itself,

$$a_{24}^{(k)} = a_{24}^{(k-1)} + 1, \;\; b_{24}^{(k)} = b_{24}^{(k-1)} + 2.$$

By contrast, dose-cycle combinations incomparable with (2,4), such as (1,2) or (3,6), are not updated through extrapolation:

$$a_{12}^{(k)} = a_{12}^{(k-1)}, b_{12}^{(k)} = b_{12}^{(k-1)},$$

and similar for $(3,6)$. After these updates, the calibration step rescales the pseudo counts to equalize the ESS across all dose-cycle combinations.

Under this ordering, the updating rule in Section 2.1 is applied with dose-cycle combinations replacing the usual two-agent combinations. That is, after each cohort is observed, pseudo-DLT information is propagated to combinations that are more toxic under the $C$ partial order, whereas pseudo-non-DLT information is propagated to combinations that are less toxic. The same ESS calibration is then used to obtain the posterior Beta parameters for all dose-cycle combinations.

In the early stage of the trial, when observed data are sparse, the proposed modification still propagates information across dose-cycle combinations through the C partial order. The calibration step then rescales the pseudo counts so that no combination has an artificially large effective sample size solely because of its position in the ordering structure. Thus, early dose recommendations are informed by the assumed ordering but are not driven by extrapolation alone.

**Modification to the Decision Rule**: In the standard two-agent c-CFBD design, escalation may occur along either dimension or both. For example, escalation from $(2,2)$ to $(2,3)$, $(3,2)$ or $(3,3)$ is possible. In CAR T-cell therapy, however, the cycle number always increases by exactly one. Thus, for a patient who has completed cycle $c-1$ at dose level $j$, the decision reduces to selecting the dose $j'$ for cycle $c$. The utility function remains as described in Section 2.1, with $p_{ij}^{(k)}$ replaced by $p_{jc}^{(k)}$. The recommended dose is therefore

$$\arg\max_j E\left[u_k\left(p_{jc}^{(k)}, p_t\right)|Data_k\right].$$

**Modification to the stopping rule**: We also modify the stopping criteria. The trial stops enrolling new patients if

$$\Pr[p_{MTD^+,1} > p_t + \delta|Data_k] > r_2,$$

where $(MTD^+, 1)$ is the next higher dose level above the current estimated MTD in cycle 1. The trial may also end early, before reaching $n_{max}$, if all currently active patients have completed their final cycle. At trial completion, a separate MTD is identified for each cycle.

The proposed modification is not intended for settings in which toxicity is constant across cycles, increases over time, or does not follow this ordering. If the assumption is incorrect, the extrapolation used in the modified c-CFBD stage may produce less reliable dose recommendations, especially in later cycles. For this reason, the proposed design should be applied only when the assumed cycle-specific toxicity pattern is supported by prior clinical knowledge and remains broadly consistent with the accumulating trial data.

## 2.3. The TWICEBEE Design

The proposed two-stage design is called TWICEBEE, which stands for TWo-stage Intra-patient Curve-frEe Bayesian dEcision-thEoretic dose escalation. The design assumes that each patient may receive up to $C$ treatment cycles and consists of two sequential stages.

**Stage 1: Accelerated Titration (AT) Stage With Fixed Dosing**

In Stage 1, an accelerated titration design is used with a prespecified starting dose chosen by the investigators and clinicians. All patients enrolled during this stage begin treatment at the same starting dose level. In each subsequent cycle,

- if no DLT is observed in any patient, the dose is escalated by one level;
- escalation continues until the patient reaches the highest dose level, $J$.

If $C > J$ and no DLT is observed at any cycle, all remaining cycles are given at dose level $J$.

Stage 1 ends when either

- a DLT is observed in any patient at any cycle or
- the number of enrolled patients reaches $n_0$, a prespecified value chosen by the investigators.

Although dosing in Stage 1 follows the AT scheme, we still track and update the hyperparameters $\alpha_{jc}$ and $\beta_{jc}$ for all dose-cycle combinations $(j, c)$ under the modified c-CFBD model. This allows an immediate and informed transition to Stage 2 when needed. This also helps stabilize the transition to the model-based stage when only limited early data are available.

**Stage 2: Modified c-CFBD Stage**

In Stage 2, the modified c-CFBD design is used to determine dose assignments for both newly enrolled patients and patients who remain on treatment. Specifically, Stage 2 serves two purposes:

- to select the starting dose for each newly enrolled patient, and
- to determine the next-cycle dose for patients who remain on treatment and did not experience a DLT in the previous cycle.

For newly enrolled patients, the starting dose is chosen from the dose-cycle combinations $(j, 1)$ using the most recent values of $\alpha_{j1}^{(k-1)}$ and $\beta_{j1}^{(k-1)}$. If Stage 2 begins before a full cohort has been treated at the current starting dose, new patients continue to be assigned to that level until the cohort is completed. After that, the modified c-CFBD design is used to recommend the starting dose for subsequent new cohorts.

For ongoing patients without a DLT in the previous cycle, the next-cycle dose is determined in two steps.

- **Step 1 (Current-Patient Decision):** For a patient currently at $(j, \mathrm{c})$, apply the modified c-CFBD design with the OSLA rule to obtain a recommended dose $(j_{curr}, c+1)$.
- **Step 2 (Previous-Patient Decision):** Identify the most recently enrolled cohort of patients who have already completed cycle $c+1$ and whose outcomes for that cycle are available. Record the dose levels they received in cycle $c+1$. For each unique dose level among them, apply the modified c-CFBD design with OSLA to obtain a recommended dose for cycle $c+1$. Let $(j_{prev}, c+1)$ be the smallest of these recommended doses.
- **Final Decision:** Assign:

$$(j^*, c+1), \qquad \text{where } j^* = \min(j_{curr}, j_{prev}).$$

  For the first enrolled patient, for whom no previous patients are available, the recommendation is simply $(j_{curr}, c+1)$.

**Special Rule for Prior-Cycle DLTs**: If a patient experiences a DLT in the prior cycle, then the next-cycle dose is determined by a simple safety rule. If the patient is not at the lowest dose level, the dose is de-escalated by one level. If the patient is already at the lowest dose level, the patient remains at that dose level in the next cycle.

**Trial Completion and Final Output**: At trial completion, we estimate the maximum tolerated dose by cycle (MTDC), that is, the MTD for each treatment cycle. These MTDC values provide a practical summary of the toxicity profile across cycles and narrow the set of dosing sequences to be considered in subsequent trial phases. An overview of the TWICEBEE decision process is shown in Figure 2.

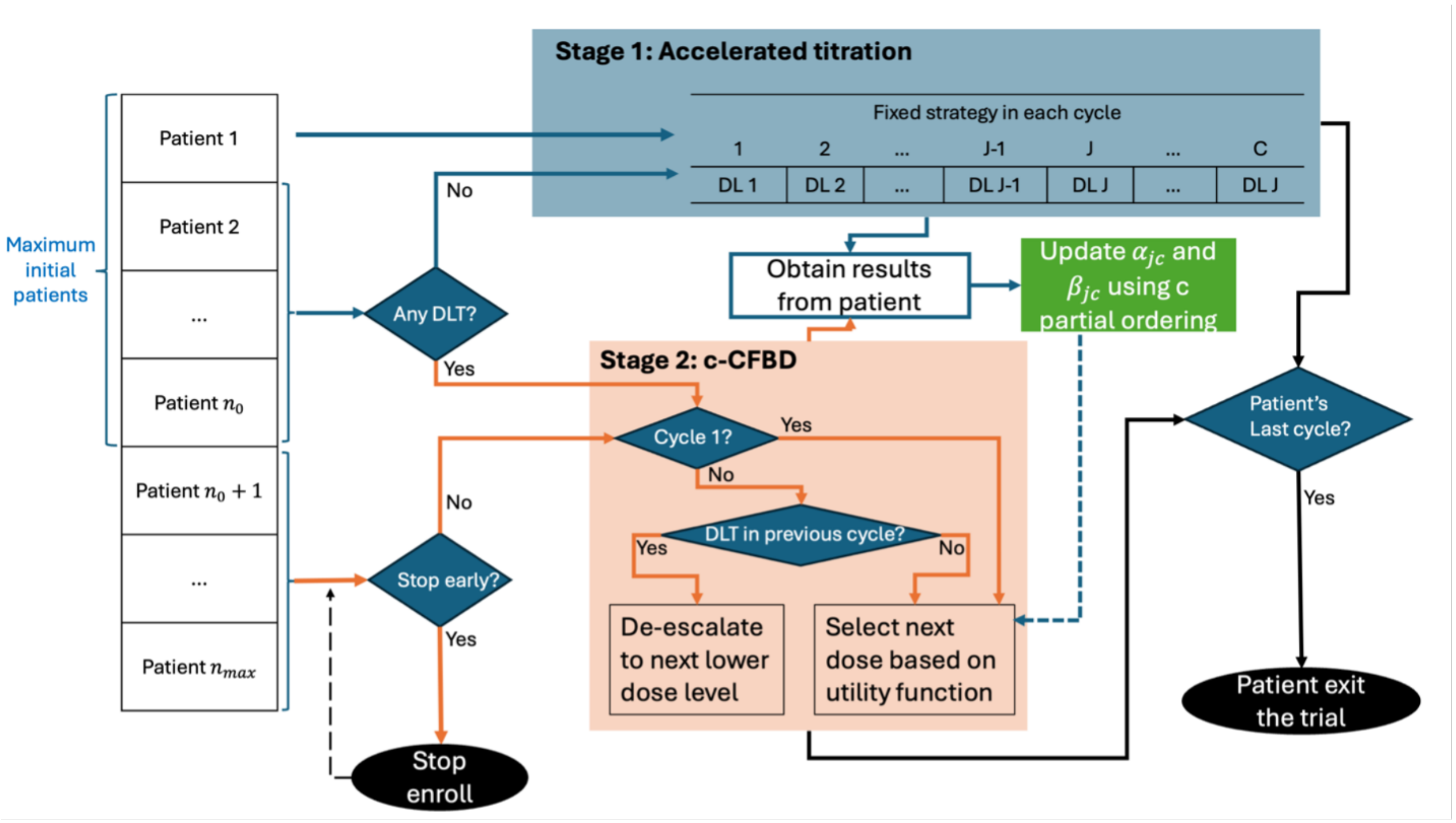

**Figure 2**: Flowchart of the TWICEBEE design.

## 3. Demonstration

We illustrate the proposed design using a trial inspired by the GD2 CAR T-cell therapy described in Section 2. In this example, the investigational therapy is a GPC2 CAR T-cell product for pediatric and young adult patients with relapsed or refractory medulloblastoma (IND 31556). The trial allows a maximum of $C = 8$ treatment cycles. We set $n_0 = 6$, $n_{max} = 12$, and target toxicity $p_t = 0.3$. Thus, the AT stage can enroll at most six patients. The total sample size is capped at 12, and at least 6 patients must be enrolled. We initialize the modified c-CFBD design as follows. Assume a prior sample size of $n_{\text{prior}} = 4$ and a prior mean of $0.2, 0.3, 0.4$ for the three dose levels in all eight cycles, this allows the construction of the priors for each dose level in each cycle of treatment. See Appendix Section A.1 for the full detail of the priors. The thresholds of the early stopping rules are chosen to be $r_1 = 0.5$, and $r_2 = 0.95$, respectively, with $\delta = 0.05$, i.e., maximum tolerated toxicity $(p_t + \delta)$ is 0.35. These settings for the early-stopping rules align with those of (Xu, et al. 2025).

A representative run of the design is shown in Appendix Section A.1. The trial switched from Stage 1 to Stage 2 after the first DLT, which occurred in cycle 3 of patient 2. Stage 2 then applied the modified c-CFBD design with the C partial ordering to recommend doses for each cycle of each patient who did not experience a prior-cycle DLT. Patients with a prior-cycle DLT were assigned according to the fixed de-escalation rule (to the next lower dose if possible, otherwise remain at the lowest dose).

In some instances, the modified c-CFBD recommended escalation to a higher dose (e.g., dose level 3) even when the patient had previously been treated at a lower level. To prevent dose "jumps," the recommendation was restricted based on the lowest relevant dose administered to either (i) the patient in the previous cycle or (ii) the most recent cohort in the current cycle of interest. For example, if the design suggested escalation to DL 3, but the patient's prior dose was DL 2 while the most recent cohort received DL 1 in the current cycle, the assigned dose would be DL 2 instead of DL 3. If safety data justify, this restriction can be relaxed to allow direct escalation when warranted.

In this run, enrollment stopped at $n = 6$ patients when

$$\Pr\left(p_{MTD^+,1} > p_t + 0.05 | Data_k\right) = 0.98 > 0.95,$$

where $MTD^+ = 3$, satisfying both the minimum sample size requirement and an early-stopping criterion. The trial stopped enrolling new patients but continued treating current patients until all completed their final cycle of treatment.

The final estimated MTDC was DL 1 for cycles 1–3, DL 2 for cycles 4–7, and DL 3 for cycle 8. Table 1 summarizes the number of patients treated and the number experiencing DLTs at each dose–cycle combination. Therefore, the RP2D is defined as DL 1 for the first three cycles, restricted to a maximum of DL 2 for cycles 4 through 7, and unrestricted for the final treatment cycle.

# 4. Operating Characteristics

## 4.1. Simulation Setup

We evaluate the operating characteristics of the proposed design and compare its performance with three alternatives:

- DIETE: The DIETE design of (Xu, Zhuo and Rasmussen 2021);
- IP-CRM: The IP-CRM design of (Guo and Liu 2025); and
- 2S-i3+3: The 2-stage with i3+3 design. This is a variant of the TWICEBEE design in which the second-stage modified c-CFBD component is replaced by the i3+3 design (Liu, Wang and Ji 2020). In this variant, the two-step procedure for determining the next-cycle dose for patients without a prior-cycle DLT uses the i3+3 method.

Specifically:

- For the **current-patient decision**, the i3+3 design is used to determine $(j_{curr}, c+1)$ based on the patient's current dose $(j, c)$ and all accumulated data up to cycle $c$.
- For the **previous-patient decision**, the i3+3 design is applied separately to each unique dose level among the previous patients, using all accumulated data up to cycle $c+1$.

Finally, isotonic regression is used to select the MTD in each cycle.

We assume a target toxicity rate of 30% $(p_t = 0.3)$. Two sets of simulation scenarios are considered, differing in the assumption whether the intra-patient toxicity events between cycles are correlated. The first set assumes independence: the true toxicity probability for each dose–cycle combination is independent across cycles, so a patient's DLT in one cycle does not affect their DLT probability in subsequent cycles. Table 2 presents ten scenarios for this cycle-independent setting, with bolded dose levels being the truth in each treatment cycle. Note that some treatment cycles in some scenarios do not have a dose level with toxicity probability being 30%. Assume a window of MTD of [0.25, 0.35], we bold the dose levels immediately below and above the true MTD (i.e., within the MTD window) in Scenarios 1 to 7. In these cases, the true MTD is the dose level between the two bolded levels. See the black solid line in the left subfigures of Figures 3 and 4. For scenarios where the true MTD is either below the minimum or above the maximum dose levels in some treatment cycles, we define the maximum attainable dose level (MAD) representing the attainable dose level that is most close to the true MTD. In these cases, the bolded dose levels represent the true MAD. In each scenario, the true toxicity probabilities satisfy

$$p_{jc} \geq p_{j'c'} \text{ if } j \geq j' \text{for fixed } c = c' \text{ or } \; c \leq c' \text{ for fixed } j = j',$$

which matches the assumptions in our CAR T-cell therapy motivating example. Scenarios 4 and 8 represent settings where toxicity strictly decreases over cycles. The remaining scenarios illustrate situations where toxicity plateaus after certain cycles.

The second set of simulations uses the same ten scenarios shown in Table 2 but introduces a positive intra-patient dependence in DLT outcomes between consecutive cycles. Specifically, Table 2 gives the baseline toxicity probability $p_{jc}$ for dose level $j$ in cycle $c$. In the dependent setting, if a patient experiences a DLT in cycle $c - 1$, then the toxicity probability in cycle $c$ is increased to

$$p'_{jc} = p_{jc} + \rho(1 - p_{jc}),$$

Whereas if no DLT is observed in cycle $c - 1$, the toxicity probability in cycle $c$ remains $p_{jc}$. We set $\rho = 0.1$ for all scenarios. Thus, the probabilities in Table 2 should be interpreted as baseline probabilities, and the effective cycle-specific toxicity profile under dependence may differ from that in the cycle-independent setting.

The maximum sample size for all methods is set to $n_{max} = 12$. For the proposed design, we set $n_0 = 6$, $\delta = 0.05, r_1 = 0.5,$ and $r_2 = 0.95$. The equivalence interval in the i3+3 design is set to $[0.25, 0.35]$. Other parameters in the DIETE and IP-CRM designs use their default values. Each design is evaluated using 1,000 simulated trials per scenario for both the independent- and correlated-cycle settings.

Comparisons focus on two metrics: **dose recommendation** and **patient allocation**. For dose recommendation, an effective design should frequently identify the correct MTDC, which is the MTD or the MAD in each cycle. For patient allocation, an effective design should assign most patients to the true MTD or MAD in each cycle while minimizing assignments above the MTD/MAD. For designs such as DIETE, patients may receive different numbers of treatment cycles; therefore, the allocation probabilities for a given cycle may not sum to one. In the dependent-cycle setting, because a prior DLT changes the toxicity probability in the next cycle, the effective cycle-specific target need not coincide exactly with the baseline MTDC implied by Table 2. For this reason, the dependent-cycle simulations are interpreted mainly as a robustness assessment of design behavior under within-patient dependence.

### 4.2. Simulation Results

We first present results from the independent-cycle simulations, where no correlation is assumed between cycles. Complete results for all ten scenarios are provided in Appendix A.2, where the full distributions of recommended and allocated dose levels are reported by cycle. In each cycle, the bolded entry corresponds to the true MTD/MAD and therefore gives the cycle-specific probability of correct recommendation or correct allocation.

Overall, the proposed TWICEBEE design accurately identified the true MTD/MADs in most treatment cycles and allocated patients appropriately. For example, in Scenarios 3–7, TWICEBEE correctly recommended the true MTD/MAD in four out of five cycles. In the remaining scenarios, it was correct in three out of five cycles. For patient allocation, TWICEBEE assigned the highest proportion of patients to the correct dose in all five cycles in Scenarios 2, 4, 5, and 7.

Compared with 2S-i3+3, TWICEBEE generally provided more accurate dose recommendations and allocated more patients to the correct dose level in Scenarios 4–10. The i3+3 variant tended to recommend and assign patients to safer, but potentially subtherapeutic, dose levels.

In comparison with the DIETE and IP-CRM designs, TWICEBEE can recommend the MTDC. In other words, it is capable of recommending different MTD/MADs for different cycles. This reflects the CAR-T therapy setting, where the MTD/MAD may vary over time. In contrast, DIETE—developed for T-cell engager therapies—reports only two MTDs (before and after an adaptation stage), resulting in identical dose recommendations across cycles 2–5. The IP-CRM design reports a single MTD/MAD for all cycles, giving the same recommendation for the entire trial.

The designs also differ in patient allocation. For DIETE and IP-CRM, the allocation proportions across the four dose levels in cycles 3–5 often do not sum to one, since patients may exit the trial before completing all cycles. In all scenarios, the DIETE design does not assign any patients to dose level 3 in cycle 5 when $J = 4$. The rationale for this is provided in Appendix A.3.

Figures 3 and 4 summarize the results graphically on the dose-level scale In Figure 3, the left subplots show, for each cycle, the Monte Carlo mean of the cycle-specific recommended dose level across the 1,000 simulated trials, while the right subplots show the standard deviation from the true MTDC, that is, the MTD or MAD in each cycle. In Figure 4, the left subplots show, for each cycle, the Monte Carlo mean of the average allocated dose level, where within each simulated trial this quantity is computed as the weighted average of the four dose levels using the patient allocation proportions in that cycle as weights. The right subplots show the corresponding standard deviation from the truth. For both figures, the black solid line denotes the true MTDC/MAD trajectory. For TWICEBEE, the expected recommended dose generally increases with cycle number, closely matching the truth in scenarios such as 5 and 6. By contrast, DIETE's curve increases and then plateaus after cycle 2, while IP-CRM's curve remains constant. The 2S-i3+3 variant consistently estimates lower dose levels than the other designs. In terms of variability, TWICEBEE achieves low variability in general for Scenarios 2 to 7 in both dose recommendation and patient allocation.

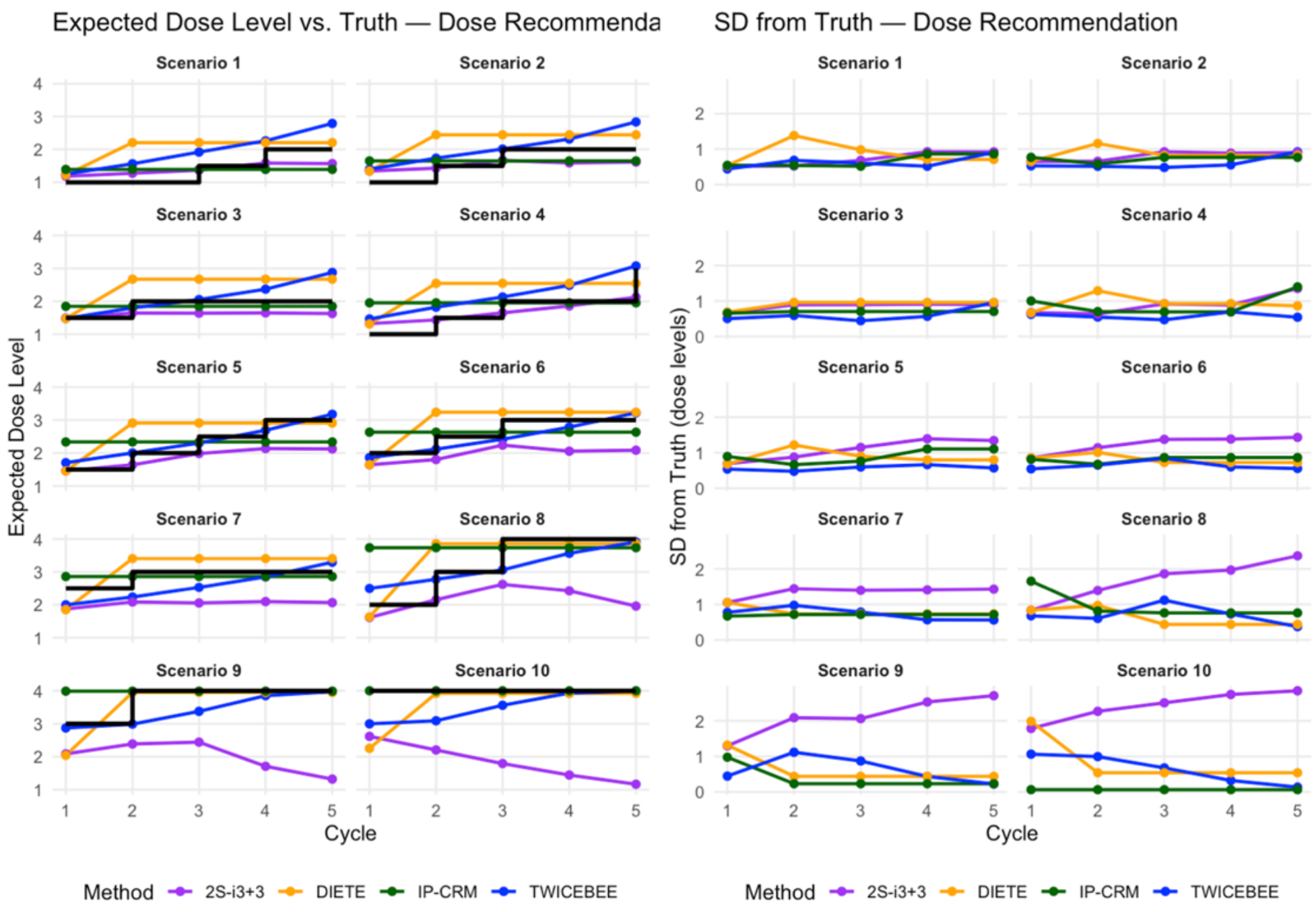

**Figure 3**: For each scenario and treatment cycle, the left panels show the Monte Carlo mean of the cycle-specific recommended dose level across 1,000 simulated trials for each method. The black solid line denotes the true MTDC/MAD trajectory. The right panels show the standard deviation of the difference between the recommended dose level and the truth.

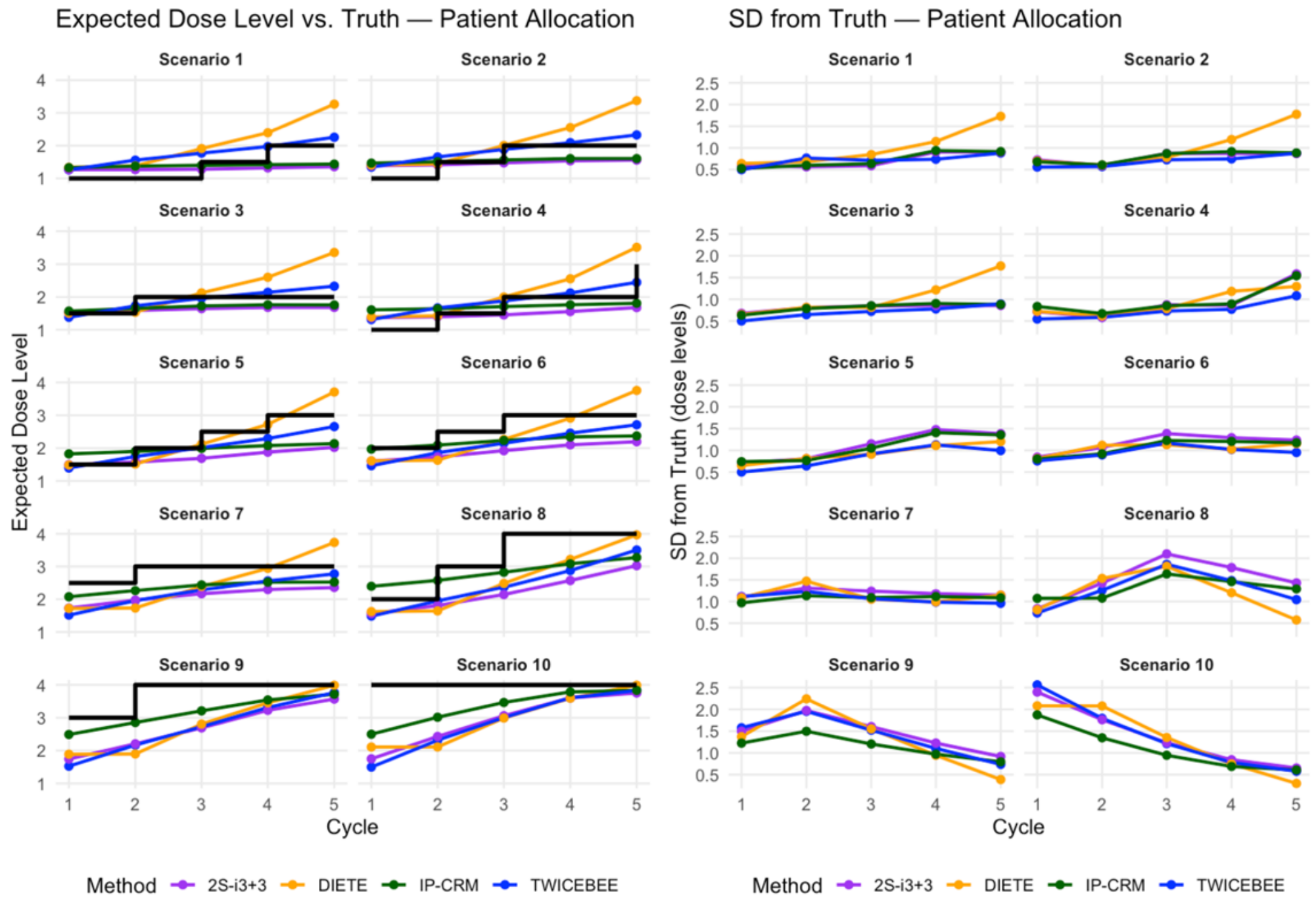

**Figure 4**: For each scenario and treatment cycle, the left panels show the Monte Carlo mean of the within-trial average allocated dose level for each method, where the average allocated dose level in a given trial and cycle is computed as the weighted average of the dose levels using the patient allocation proportions in that cycle as weights. The black solid line denotes the true MTDC/MAD trajectory. The right panels show the standard deviation of the difference between the allocated dose level and the truth.

For patient allocation, the i3+3 variant is the least aggressive overall. DIETE is generally the most aggressive in Scenarios 1–7, while IP-CRM is most aggressive in Scenarios 8–10. TWICEBEE is moderately aggressive, aligning closely with the truth in Scenarios 2–7. Overall, TWICEBEE achieves MTDC recommendations that are close to the truth and allocates patients appropriately across treatment cycles.

Lastly, as shown in Figure 5, the TWICEBEE design has an additional advantage: it uses the smallest expected sample size among the designs considered. This is a property inherited from the c-CFBD design (Xu, et al. 2025).

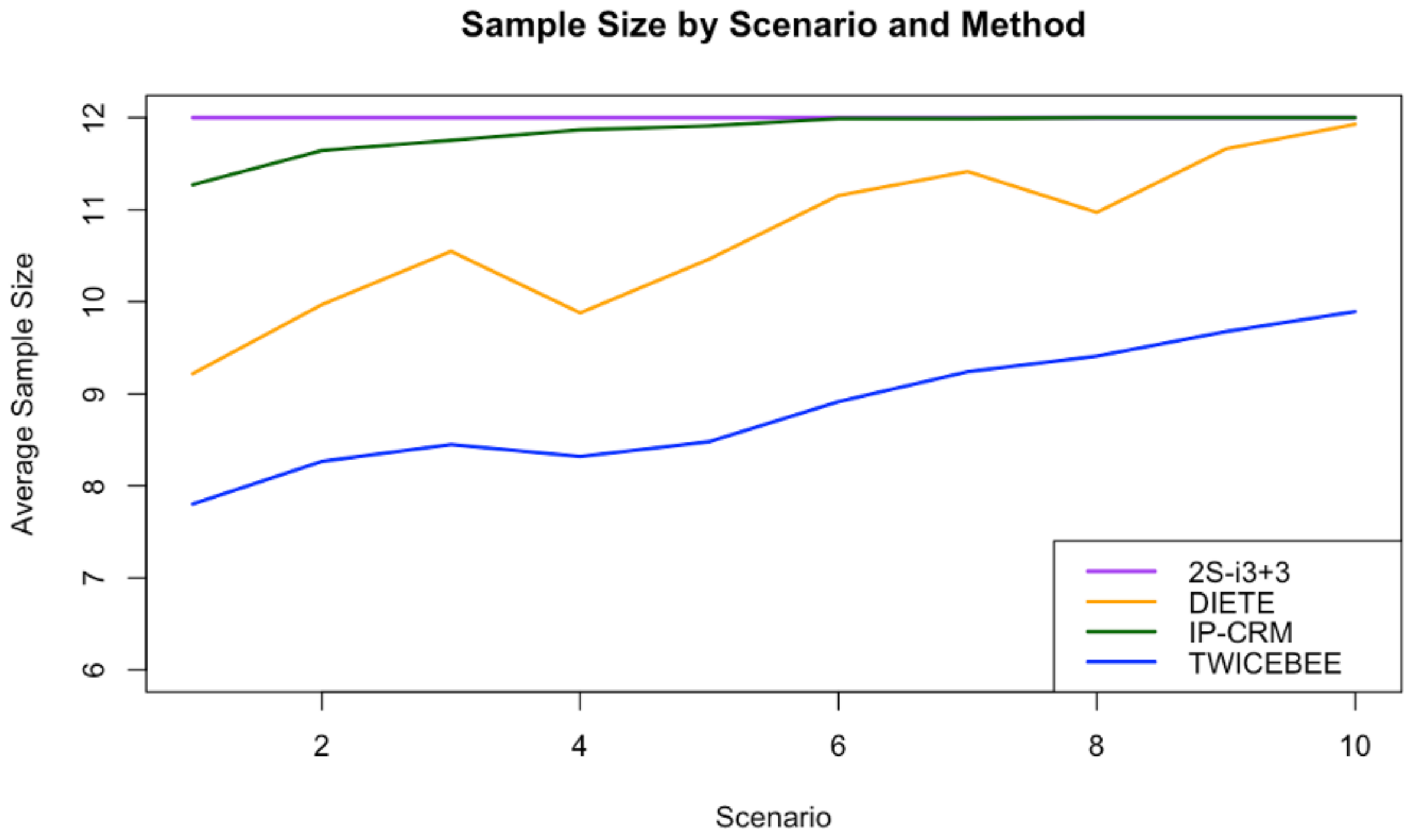

**Figure 5**: Average sample size of the methods under comparison across the ten simulation scenarios.

For the second set of simulations, results across the ten scenarios were qualitatively similar to those in the cycle-independent setting (see Appendix A.4). The proposed design again yielded the lowest expected sample size. At the same time, all methods showed a general shift toward recommending lower doses and allocating more patients to lower doses. This pattern is consistent with the dependent simulation setup, under which a DLT in the previous cycle increases the probability of DLT in the current cycle. Because the effective cycle-specific toxicity profile under dependence need not be the same as that implied by the baseline probabilities in Table 2, these results are best interpreted as a robustness assessment rather than a direct like-for-like comparison with the independent-case setting.

## 5. Discussion

In this work, we proposed TWICEBEE, a two-stage intra-patient dose-escalation design for multi-cycle dose finding, which is tailored for immunotherapies that toxicity for a given dose decreases over cycles. Our design is motivated by intracranial CAR T-cell therapy. However, it is also applicable to a broader class of multi-cycle immune-based therapies, including repeated or locoregional CAR-T platforms (Brown, et al. 2024, Vitanza, et al. 2025, Vu, et al. 2026) as well as bispecific T-cell engager (Kim, et al. 2025) and bispecific antibody (Ahn, et al. 2023) regimens, where immune-related toxicity is often most pronounced in the initial administrations and may decrease over later cycles, thereby creating an opportunity for higher and potentially more effective doses at later cycles. These examples highlight the key methodological feature that motivates TWICEBEE: dose finding must account for both dose level and treatment cycle, rather than targeting a single MTD for the entire treatment course. This broader perspective also suggests that the underlying design principle may be useful beyond the present setting. For example, in therapies (Mader, et al. 2023) where toxicity accumulates over time, the ordering along the cycle dimension could be reversed accordingly. To address the motivating setting of nonincreasing toxicity across cycles, TWICEBEE begins with an accelerated titration (AT) stage to rapidly explore dose levels, and after the first observed DLT, transitions to a c-CFBD-based design to improve safety. Simulation studies showed that TWICEBEE is straightforward to implement and has favorable operating characteristics in this setting.

Intrapatient dose escalation has not been commonly used in dose-finding trials, but recent FDA dose-optimization guidance states that strategies such as priming doses and intra-patient dose escalation or de-escalation should be explored when appropriate. The same guidance also recommends comparing multiple dosages in trials designed to assess antitumor activity, safety, and tolerability, and notes that model-informed approaches may help identify the dosages to be studied. Although the guidance does not specifically address cellular and gene therapy products, it notes that some of its recommendations may still apply to those modalities. In this sense, TWICEBEE is well aligned with the broader goal of prospective dosage optimization in oncology and may help identify a small set of cycle-specific dose schedules for further study in settings where repeated dosing within a patient is clinically meaningful. At the same time, the simulation comparisons should be interpreted considering the study objective. TWICEBEE targets a sequence of cycle-specific MTDs, whereas methods such as DIETE and IP-CRM aim at a more limited summary of toxicity over time. Thus, TWICEBEE has a natural advantage when the true MTD varies across cycles, and the comparisons should be understood as reflecting performance in the motivating setting rather than universal superiority across all intra-patient dose-finding problems.

The design has several limitations. The extrapolation used in the c-CFBD stage may overestimate toxicity at higher dose levels in cycle 1 and underestimate toxicity at lower dose levels in the final cycle. In addition, the current framework uses toxicity outcomes alone, which may be suboptimal in settings such as CAR T-cell therapy where a monotone toxicity–efficacy relationship may not hold. Finally, TWICEBEE is not intended for all intra-patient dose-escalation problems and should be used only when the assumption of nonincreasing toxicity across cycles at a fixed dose is clinically justified. Future work will extend the method to incorporate efficacy outcomes and more flexible toxicity patterns.

## Acknowledgement

We thank Dr. Yuan Ji for insightful discussions on this topic. DB's effort is partially supported by CIRM and National Institutes of Health (NIH) with project number R01 CA263500-01 and P01 CA049605. YL's effort is partially supported by grants from NIH 3P30CA12443517S2, P01 CA049605, and P01 CA257907. JT and RL are supported by P01 CA049605. This work was supported by grant funding from NIH2P30CA124435-16, R35CA283888, 1UG3CA297747-01 and C.L.M. is a member of the Parker Institute for Cancer Immunotherapy, which supports the Stanford University Cancer Immunotherapy Program and is a Weill West Coast Cancer Hub Investigator.

# Tables

| Dose levels | Cycles | | | | | | | | Overall |
|---|---|---|---|---|---|---|---|---|---|
| | 1 | 2 | 3 | 4 | 5 | 6 | 7 | 8 | |
| DL 1 | 0/4 | 0/1 | 0/1 | 0/0 | 1/1 | 0/2 | 1/2 | 0/1 | 2/12 |
| DL 2 | 1/2 | 1/5 | 0/2 | 1/5 | 1/3 | 2/4 | 0/3 | 2/2 | 8/26 |
| DL 3 | 0/0 | 0/0 | 2/3 | 0/1 | 2/2 | 0/0 | 1/1 | 1/3 | 6/10 |
| Overall | 1/6 | 1/6 | 2/6 | 1/6 | 4/6 | 2/6 | 2/6 | 3/6 | 16/48 |

**Table 1**: Summary of dose assignments and DLT outcomes in the demonstration trial. Entries are given as number of DLTs / number of patients treated.

| Cycle | | | | | | Cycle | | | | |
|---|---|---|---|---|---|---|---|---|---|---|
| | **1** | **2** | **3** | **4** | **5** | | **1** | **2** | **3** | **4** | **5** |
| | **Scenario 1** | | | | | | **Scenario 2** | | | | |
| DL 1 | **0.35** | **0.30** | **0.25** | 0.20 | 0.20 | DL 1 | **0.30** | **0.25** | 0.20 | 0.20 | 0.20 |
| DL 2 | 0.45 | 0.40 | **0.35** | **0.30** | **0.30** | DL 2 | 0.40 | **0.35** | **0.30** | **0.30** | **0.30** |
| DL 3 | 0.55 | 0.50 | 0.45 | 0.40 | 0.40 | DL 3 | 0.50 | 0.45 | 0.40 | 0.40 | 0.40 |
| DL 4 | 0.65 | 0.60 | 0.55 | 0.50 | 0.50 | DL 4 | 0.60 | 0.55 | 0.50 | 0.50 | 0.50 |
| | **Scenario 3** | | | | | | **Scenario 4** | | | | |
| DL 1 | **0.25** | 0.20 | 0.20 | 0.20 | 0.20 | DL 1 | **0.30** | **0.25** | 0.20 | 0.15 | 0.10 |
| DL 2 | **0.35** | **0.30** | **0.30** | **0.30** | **0.30** | DL 2 | 0.40 | **0.35** | **0.30** | **0.25** | 0.20 |
| DL 3 | 0.45 | 0.40 | 0.40 | 0.40 | 0.40 | DL 3 | 0.50 | 0.45 | 0.40 | 0.35 | **0.30** |
| DL 4 | 0.55 | 0.50 | 0.50 | 0.50 | 0.50 | DL 4 | 0.60 | 0.55 | 0.50 | 0.45 | 0.40 |
| | **Scenario 5** | | | | | | **Scenario 6** | | | | |
| DL 1 | **0.25** | 0.20 | 0.15 | 0.10 | 0.10 | DL 1 | 0.20 | 0.15 | 0.10 | 0.10 | 0.10 |
| DL 2 | **0.35** | **0.30** | **0.25** | 0.20 | 0.20 | DL 2 | **0.30** | **0.25** | 0.20 | 0.20 | 0.20 |
| DL 3 | 0.45 | 0.40 | **0.35** | **0.30** | **0.30** | DL 3 | 0.40 | **0.35** | **0.30** | **0.30** | **0.30** |
| DL 4 | 0.55 | 0.50 | 0.45 | 0.40 | 0.40 | DL 4 | 0.50 | 0.45 | 0.40 | 0.40 | 0.40 |
| | **Scenario 7** | | | | | | **Scenario 8** | | | | |
| DL 1 | 0.15 | 0.10 | 0.10 | 0.10 | 0.10 | DL 1 | 0.20 | 0.10 | 0.05 | 0.03 | 0.01 |
| DL 2 | **0.25** | 0.20 | 0.20 | 0.20 | 0.20 | DL 2 | **0.30** | 0.20 | 0.10 | 0.05 | 0.03 |
| DL 3 | **0.35** | **0.30** | **0.30** | **0.30** | **0.30** | DL 3 | 0.40 | **0.30** | 0.20 | 0.10 | 0.05 |
| DL 4 | 0.45 | 0.40 | 0.40 | 0.40 | 0.40 | DL 4 | 0.50 | 0.40 | **0.30** | **0.20** | **0.10** |
| | **Scenario 9** | | | | | | **Scenario 10** | | | | |
| DL 1 | 0.10 | 0.05 | 0.03 | 0.01 | 0.01 | DL 1 | 0.05 | 0.03 | 0.01 | 0.01 | 0.01 |

| | | | | | | | | | | |
|---|---|---|---|---|---|---|---|---|---|---|
| DL 2 | 0.20 | 0.10 | 0.05 | 0.03 | 0.03 | DL 2 | 0.10 | 0.05 | 0.03 | 0.03 | 0.03 |
| DL 3 | **0.30** | 0.20 | 0.10 | 0.05 | 0.05 | DL 3 | 0.20 | 0.10 | 0.05 | 0.05 | 0.05 |
| DL 4 | 0.40 | **0.30** | **0.20** | **0.10** | **0.10** | DL 4 | **0.30** | **0.20** | **0.10** | **0.10** | **0.10** |

**Table 2**: Simulation Truths for Scenarios 1 – 10. Each panel displays the true toxicity probability for each dose level (DL) across five cycles, assuming $J = 4$ dose levels. Bold values indicate the true MTD/MAD for each cycle within each scenario.

# Appendix

## A.1 Example Run of Proposed Design

We provide an example run of the proposed TWICEBEE design in Section 3. Table A.1 shows Stage 1, the AT stage. Below shows the prior we used in this demonstration:

| DL/Cyc | 1 | 2 | 3 | 4 | 5 | 6 | 7 | 8 |
|---|---|---|---|---|---|---|---|---|
| DL 1 | Beta(0.8,3.2) | Beta(0.8,3.2) | Beta(0.8,3.2) | Beta(0.8,3.2) | Beta(0.8,3.2) | Beta(0.8,3.2) | Beta(0.8,3.2) | Beta(0.8,3.2) |
| DL 2 | Beta(1.2,2.8) | Beta(1.2,2.8) | Beta(1.2,2.8) | Beta(1.2,2.8) | Beta(1.2,2.8) | Beta(1.2,2.8) | Beta(1.2,2.8) | Beta(1.2,2.8) |
| DL 3 | Beta(1.6,2.4) | Beta(1.6,2.4) | Beta(1.6,2.4) | Beta(1.6,2.4) | Beta(1.6,2.4) | Beta(1.6,2.4) | Beta(1.6,2.4) | Beta(1.6,2.4) |

Tables A.2 to A.6 show the second stage, the c-CFBD stage. In particular, Table A.2 presents a step-by-step decision process. Tables A.3 to A.6 display the decisions and observed toxicities for the remainder of the trial. Table A.7 summarizes the patient-treatment path.

| Patient ID | Enroll time | Cycles | | | | | | | | | | | | | | | |
|---|---|---|---|---|---|---|---|---|---|---|---|---|---|---|---|---|---|
| | | 1 | | 2 | | 3 | | 4 | | 5 | | 6 | | 7 | | 8 | |
| **Stage 1: Accelerated Titration Stage** | | | | | | | | | | | | | | | | | |
| 1 | Week 1 | DL 1 | | | | | | | | | | | | | | | |
| 1 | Week 4 (28 days from Week 1) | DL 1 | 0 | DL 2 | | | | | | | | | | | | | |
| 2 | | DL 1 | | | | | | | | | | | | | | | |
| 1 | Week 8 (28 days from Week 4) | DL 1 | 0 | DL 2 | 0 | DL 3 | | | | | | | | | | | |
| 2 | | DL 1 | 0 | DL 2 | | | | | | | | | | | | | |
| 3 | | DL 1 | | | | | | | | | | | | | | | |
| 1 | Week 12 (28 days from Week 12) | DL 1 | 0 | DL 2 | 0 | DL 3 | 0 | DL 3 | | | | | | | | | |
| 2 | | DL 1 | 0 | DL 2 | 0 | DL 3 | | | | | | | | | | | |
| 3 | | DL 1 | 0 | DL 2 | | | | | | | | | | | | | |
| 4 | | DL 1 | | | | | | | | | | | | | | | |

**Table A.1**: An example of the accelerated titration stage of the proposed design. Here, three dose levels are available, and each patient can receive up to eight cycles. In this stage, all patients start

at dose level 1 in cycle 1, escalate to dose level 2 in cycle 2, and then escalate to dose level 3 in cycle 3. For the remaining cycles, patients stay at dose level 3 if no DLTs are observed.

| Patient ID | Enroll time | Cycles 1 | | 2 | | 3 | | 4 | | 5 | | 6 | | 7 | | 8 | |
|---|---|---|---|---|---|---|---|---|---|---|---|---|---|---|---|---|---|
| 1 | Week 12 (28 days from Week 12) | DL 1 | 0 | DL 2 | 0 | DL 3 | 0 | DL 3 | 0 | | | | | | | | |
| 2 | | DL 1 | 0 | DL 2 | 0 | DL 3 | 1 | | | | | | | | | | |
| 3 | | DL 1 | 0 | DL 2 | 0 | | | | | | | | | | | | |
| 4 | | DL 1 | 0 | | | | | | | | | | | | | | |
| **Stage 2: c-CFBD Stage** | | | | | | | | | | | | | | | | | |
| 1 | Week 16 (28 days from Week 16) | DL 1 | 0 | DL 2 | 0 | DL 3 | 0 | DL 3 | 0 →(S) | DL 3 | | | | | | | |
| 2 | | DL 1 | 0 | DL 2 | 0 | DL 3 ↓(S) | 1 →(D) | DL 2 | | | | | | | | | |
| 3 | | DL 1 | 0 | DL 2 ↓(S) | 0 →(E) | DL 3 | | | | | | | | | | | |
| 4 | | DL 1 ↓(E) | 0 →(E) | DL 2 | | | | | | | | | | | | | |
| 5 | | DL 2 | | | | | | | | | | | | | | | |

| DL\Cyc | 1 | 2 | 3 | 4 | 5 | 6 | 7 | 8 |
|---|---|---|---|---|---|---|---|---|
| 1 | Beta(0.8,7.2) | Beta(0.8,10.2) | Beta(0.8,11.2) | Beta(0.8,12.2) | Beta(0.8,12.2) | Beta(0.8,12.2) | Beta(0.8,12.2) | Beta(0.8,12.2) |
| 2 | Beta(1.2,2.8) | Beta(1.2,5.8) | Beta(1.2,6.8) | Beta(1.2,7.8) | Beta(1.2,7.8) | Beta(1.2,7.8) | Beta(1.2,7.8) | Beta(1.2,7.8) |
| 3 | Beta(2.6,2.4) | Beta(2.6,2.4) | Beta(2.6,3.4) | Beta(1.6,4.4) | Beta(1.6,4.4) | Beta(1.6,4.4) | Beta(1.6,4.4) | Beta(1.6,4.4) |

$\begin{cases} \text{DL 2, C2} \xrightarrow{E} \text{DL 3} \\ \text{DL 3, C3} \xrightarrow{S} \text{DL 3} \end{cases}$ min → DL 3

$\begin{cases} \text{DL 1, C1} \xrightarrow{E} \text{DL 2} \\ \text{DL 2, C2} \xrightarrow{S} \text{DL 2} \end{cases}$ min → DL 2

Utilities at cycle 5: (−0.241 [DL 1], −0.182 [DL 2], **−0.122** [**DL 3**])

De-escalate due to previous DLT

Utilities at cycle 3: (−0.237 [DL 1], −0.170 [DL 2], **−0.168** [**DL 3**])

Utilities at cycle 2: (−0.231 [DL 1], **−0.157** [**DL 2**], −0.232 [DL 3])

Utilities at cycle 1: (−0.208 [DL 1], **−0.120** [**DL 2**], −0.232 [DL 3])

**Table A.2**: This table first shows the switch from the accelerated titration stage to the c-CFBD stage. The red 1 in Cycle 3 for patient 2 is the 1st observed DLT that triggers the transition from Stage 1 to 2. Note that the priors are updated even in the AT stage. For example, the non-DLT observation of patient 1 in Cycle 1 adds 1 to the $b_{jc}^{(0)} = 3.2$ for $j = 1$ and $c = 1, \dots, 8$. The small table shows the updated prior when we switch from the AT stage to the c-CFBD stage. It then presents a step-by-step decision process in the c-CFBD stage. Decisions are shown for each current patient's subsequent cycle and for the first cycle of each newly enrolled patient. The cohort size is assumed to be one.

| Patient ID | Enroll time | Cycles | | | | | | | | | | | | | | | |
|---|---|---|---|---|---|---|---|---|---|---|---|---|---|---|---|---|---|
| | | 1 | | 2 | | 3 | | 4 | | 5 | | 6 | | 7 | | 8 | |
| 1 | Week 20 (28 days from Week 16) | DL 1 | 0 | DL 2 | 0 | DL 3 | 0 | DL 3 | 0 | DL 3 | 1 | DL 2 | 1 | | | | |
| 2 | | DL 1 | 0 | DL 2 | 0 | DL 3 | 1 | DL 2 | 0 | DL 3 | 1 | | | | | | |
| 3 | | DL 1 | 0 | DL 2 | 0 | DL 3 | 1 | DL 2 | 0 | | | | | | | | |
| 4 | | DL 1 | 0 | DL 2 | 1 | DL 1 | 0 | | | | | | | | | | |
| 5 | | DL 2 | 0 | DL 2 | 0 | | | | | | | | | | | | |
| 6 | | DL 2 | 1 | | | | | | | | | | | | | | |
| **Stop enrolling new patients** | | | | | | | | | | | | | | | | | |
| 1 | Week 24 (28 days from Week 20) | DL 1 | 0 | DL 2 | 0 | DL 3 | 0 | DL 3 | 0 | DL 3 | 1 | DL 2 | 1 | DL 1 | 0 | | |
| 2 | | DL 1 | 0 | DL 2 | 0 | DL 3 | 1 | DL 2 | 0 | DL 3 | 1 | DL 2 | 0 | | | | |
| 3 | | DL 1 | 0 | DL 2 | 0 | DL 3 | 1 | DL 2 | 0 | DL 2 | 1 | | | | | | |
| 4 | | DL 1 | 0 | DL 2 | 1 | DL 1 | 0 | DL 2 | 0 | | | | | | | | |
| 5 | | DL 2 | 0 | DL 2 | 0 | DL 2 | 0 | | | | | | | | | | |
| 6 | | DL 2 | 1 | DL 1 | 0 | | | | | | | | | | | | |

**Table A.3**: The dosing decisions and the observed DLTs for each patient. We stop enroll new patients because there are already 6 patients, and $\Pr\left[p_{3,1} > p_t + \delta | Data\right] > 0.95$.

| Patient ID | Enroll time | Cycles | | | | | | | | | | | | | | | |
|---|---|---|---|---|---|---|---|---|---|---|---|---|---|---|---|---|---|
| | | 1 | | 2 | | 3 | | 4 | | 5 | | 6 | | 7 | | 8 | |
| **Stop enrolling new patients** | | | | | | | | | | | | | | | | | |
| 1 | Week 28 (28 days from Week 24 | DL 1 | 0 | DL 2 | 0 | DL 3 | 0 | DL 3 | 0 | DL 3 | 1 | DL 2 | 1 | DL 1 | 0 | DL 2 | 1 |
| 2 | | DL 1 | 0 | DL 2 | 0 | DL 3 | 1 | DL 2 | 0 | DL 3 | 1 | DL 2 | 0 | DL 3 | 1 | | |
| 3 | | DL 1 | 0 | DL 2 | 0 | DL 3 | 1 | DL 2 | 0 | DL 2 | 1 | DL 1 | 0 | | | | |
| 4 | | DL 1 | 0 | DL 2 | 1 | DL 1 | 0 | DL 2 | 0 | DL 2 | 0 | | | | | | |
| 5 | | DL 2 | 0 | DL 2 | 0 | DL 2 | 0 | DL 2 | 1 | | | | | | | | |
| 6 | | DL 2 | 1 | DL 1 | 0 | DL 2 | 0 | | | | | | | | | | |
| 1 | Week 32 (28 days from Week 28 | DL 1 | 0 | DL 2 | 0 | DL 3 | 0 | DL 3 | 0 | DL 3 | 1 | DL 2 | 1 | DL 1 | 0 | DL 2 | 1 |
| 2 | | DL 1 | 0 | DL 2 | 0 | DL 3 | 1 | DL 2 | 0 | DL 3 | 1 | DL 2 | 0 | DL 3 | 1 | DL 3 | 0 |
| 3 | | DL 1 | 0 | DL 2 | 0 | DL 3 | 1 | DL 2 | 0 | DL 2 | 1 | DL 1 | 0 | DL 2 | 0 | | |
| 4 | | DL 1 | 0 | DL 2 | 1 | DL 1 | 0 | DL 2 | 0 | DL 2 | 0 | DL 2 | 1 | | | | |
| 5 | | DL 2 | 0 | DL 2 | 0 | DL 2 | 0 | DL 2 | 1 | DL 1 | 1 | | | | | | |
| 6 | | DL 2 | 1 | DL 1 | 0 | DL 2 | 0 | DL 2 | 0 | | | | | | | | |

**Table A.4**: The table shows the dosing decision and observed DLT for each patient. A grey row indicates that the patient has completed their final treatment cycle and exited the trial.

| Patient ID | Enroll time | Cycles | | | | | | | | | | | | | | | |
|---|---|---|---|---|---|---|---|---|---|---|---|---|---|---|---|---|---|
| | | 1 | | 2 | | 3 | | 4 | | 5 | | 6 | | 7 | | 8 | |
| **Stop enrolling new patients** | | | | | | | | | | | | | | | | | |
| 1 | Week 36 (28 days from Week 32 | DL 1 | 0 | DL 2 | 0 | DL 3 | 0 | DL 3 | 0 | DL 3 | 1 | DL 2 | 1 | DL 1 | 0 | DL 2 | 1 |
| 2 | | DL 1 | 0 | DL 2 | 0 | DL 3 | 1 | DL 2 | 0 | DL 3 | 1 | DL 2 | 0 | DL 3 | 1 | DL 3 | 0 |
| 3 | | DL 1 | 0 | DL 2 | 0 | DL 3 | 1 | DL 2 | 0 | DL 2 | 1 | DL 1 | 0 | DL 2 | 0 | DL 3 | 1 |
| 4 | | DL 1 | 0 | DL 2 | 1 | DL 1 | 0 | DL 2 | 0 | DL 2 | 0 | DL 2 | 1 | DL 1 | 1 | | |
| 5 | | DL 2 | 0 | DL 2 | 0 | DL 2 | 0 | DL 2 | 1 | DL 1 | 1 | DL 1 | 0 | | | | |
| 6 | | DL 2 | 1 | DL 1 | 0 | DL 2 | 0 | DL 2 | 0 | DL 2 | 0 | | | | | | |
| 1 | Week 40 (28 days from Week 36 | DL 1 | 0 | DL 2 | 0 | DL 3 | 0 | DL 3 | 0 | DL 3 | 1 | DL 2 | 1 | DL 1 | 0 | DL 2 | 1 |
| 2 | | DL 1 | 0 | DL 2 | 0 | DL 3 | 1 | DL 2 | 0 | DL 3 | 1 | DL 2 | 0 | DL 3 | 1 | DL 3 | 0 |
| 3 | | DL 1 | 0 | DL 2 | 0 | DL 3 | 1 | DL 2 | 0 | DL 2 | 1 | DL 1 | 0 | DL 2 | 0 | DL 3 | 1 |
| 4 | | DL 1 | 0 | DL 2 | 1 | DL 1 | 0 | DL 2 | 0 | DL 2 | 0 | DL 2 | 1 | DL 1 | 1 | DL 1 | 0 |
| 5 | | DL 2 | 0 | DL 2 | 0 | DL 2 | 0 | DL 2 | 1 | DL 1 | 1 | DL 1 | 0 | DL 2 | 0 | | |
| 6 | | DL 2 | 1 | DL 1 | 0 | DL 2 | 0 | DL 2 | 0 | DL 2 | 0 | DL 2 | 0 | | | | |

**Table A.5**: The table shows the dosing decision and observed DLT for each patient. A grey row indicates that the patient has completed their final treatment cycle and exited the trial.

| Patient ID | Enroll time | Cycles | | | | | | | | | | | | | | | |
|---|---|---|---|---|---|---|---|---|---|---|---|---|---|---|---|---|---|
| | | 1 | | 2 | | 3 | | 4 | | 5 | | 6 | | 7 | | 8 | |
| **Stop enrolling new patients** | | | | | | | | | | | | | | | | | |
| 1 | Week 44 (28 days from Week 40) | DL 1 | 0 | DL 2 | 0 | DL 3 | 0 | DL 3 | 0 | DL 3 | 1 | DL 2 | 1 | DL 1 | 0 | DL 2 | 1 |
| 2 | | DL 1 | 0 | DL 2 | 0 | DL 3 | 1 | DL 2 | 0 | DL 3 | 1 | DL 2 | 0 | DL 3 | 1 | DL 3 | 0 |
| 3 | | DL 1 | 0 | DL 2 | 0 | DL 3 | 1 | DL 2 | 0 | DL 2 | 1 | DL 1 | 0 | DL 2 | 0 | DL 3 | 1 |
| 4 | | DL 1 | 0 | DL 2 | 1 | DL 1 | 0 | DL 2 | 0 | DL 2 | 0 | DL 2 | 1 | DL 1 | 1 | DL 1 | 0 |
| 5 | | DL 2 | 0 | DL 2 | 0 | DL 2 | 0 | DL 2 | 1 | DL 1 | 1 | DL 1 | 0 | DL 2 | 0 | DL 2 | 1 |
| 6 | | DL 2 | 1 | DL 1 | 0 | DL 2 | 0 | DL 2 | 0 | DL 2 | 0 | DL 2 | 0 | DL 2 | 0 | | |
| 1 | Week 48 (28 days from Week 44) | DL 1 | 0 | DL 2 | 0 | DL 3 | 0 | DL 3 | 0 | DL 3 | 1 | DL 2 | 1 | DL 1 | 0 | DL 2 | 1 |
| 2 | | DL 1 | 0 | DL 2 | 0 | DL 3 | 1 | DL 2 | 0 | DL 3 | 1 | DL 2 | 0 | DL 3 | 1 | DL 3 | 0 |
| 3 | | DL 1 | 0 | DL 2 | 0 | DL 3 | 1 | DL 2 | 0 | DL 2 | 1 | DL 1 | 0 | DL 2 | 0 | DL 3 | 1 |
| 4 | | DL 1 | 0 | DL 2 | 1 | DL 1 | 0 | DL 2 | 0 | DL 2 | 0 | DL 2 | 1 | DL 1 | 1 | DL 1 | 0 |
| 5 | | DL 2 | 0 | DL 2 | 0 | DL 2 | 0 | DL 2 | 1 | DL 1 | 1 | DL 1 | 0 | DL 2 | 0 | DL 2 | 1 |
| 6 | | DL 2 | 1 | DL 1 | 0 | DL 2 | 0 | DL 2 | 0 | DL 2 | 0 | DL 2 | 0 | DL 2 | 0 | DL 3 | 0 |
| **End of the trial** | | | | | | | | | | | | | | | | | |

**Table A.6**: The table shows the dosing decision and observed DLT for each patient. A grey row indicates that the patient has completed their final treatment cycle and exited the trial.

| ID | Enroll Time | Cycle 1 | | Cycle 2 | | Cycle 3 | | Cycle 4 | | Cycle 5 | | Cycle 6 | | Cycle 7 | | Cycle 8 | | Notes |
|---|---|---|---|---|---|---|---|---|---|---|---|---|---|---|---|---|---|---|
| | | DL | DLT | DL | DLT | DL | DLT | DL | DLT | DL | DLT | DL | DLT | DL | DLT | DL | DLT | |
| Stage 1: AT stage | | | | | | | | | | | | | | | | | | |
| 1 | W 1 | 1 | 0 | 2 | 0 | 3 | 0 | 3 | 0 | | | | | | | | | No DLT |
| 2 | W 4 | 1 | 0 | 2 | 0 | 3 | 1 | | | | | | | | | | | DLT at C3 |
| 3 | W 8 | 1 | 0 | 2 | 0 | | | | | | | | | | | | | Ongoing |
| 4 | W 12 | 1 | 0 | | | | | | | | | | | | | | | New pat |
| Stage 2: c-CFBD stage | | | | | | | | | | | | | | | | | | |
| 5 | W 16 | 2 | 0 | 2 | 0 | | | | | | | | | | | | | Stage 2 |
| 6 | W 20 | 2 | 1 | | | | | | | | | | | | | | | Stop enroll |
| Enrollment stopped; follow-up of active patients continues until all patients done their last cycle of treatment | | | | | | | | | | | | | | | | | | |
| 1 | W 1 | 1 | 0 | 2 | 0 | 3 | 0 | 3 | 0 | 3 | 1 | 2 | 1 | 1 | 0 | 2 | 1 | Completed |
| 2 | W 4 | 1 | 0 | 2 | 0 | 3 | 1 | 2 | 0 | 3 | 1 | 2 | 0 | 3 | 1 | 3 | 0 | Completed |
| 3 | W 8 | 1 | 0 | 2 | 0 | 3 | 1 | 2 | 0 | 2 | 1 | 1 | 0 | 2 | 0 | 3 | 1 | Completed |
| 4 | W 12 | 1 | 0 | 2 | 1 | 1 | 0 | 2 | 0 | 2 | 0 | 2 | 1 | 1 | 1 | 1 | 0 | Completed |
| 5 | W 16 | 2 | 0 | 2 | 0 | 2 | 0 | 2 | 1 | 1 | 1 | 1 | 0 | 2 | 0 | 2 | 1 | Completed |
| 6 | W20 | 2 | 1 | 1 | 0 | 2 | 0 | 2 | 0 | 2 | 0 | 2 | 0 | 2 | 0 | 3 | 0 | Completed |

**Table A.7**: Illustration of patient-level treatment paths, dose assignments, and DLT outcomes under the TWICEBEE design. ID refers to patient ID, "W" refers to enrollment week, and DL refers to dose level. DLT indicates whether a dose-limiting toxicity occurred in that cycle (1 = yes, 0 = no). Blank cells indicate cycles not yet received or not observed. Patients may appear in more than one section of the table because follow-up continues after stage transition or after enrollment is stopped.

## A.2 Additional Simulation Results for Independent Scenarios

| Method | Dose Level | Cycle 1 | Cycle 2 | Cycle 3 | Cycle 4 | Cycle 5 | Cycle 1 | Cycle 2 | Cycle 3 | Cycle 4 | Cycle 5 |
|---|---|---|---|---|---|---|---|---|---|---|---|
| | | **Dose Recommendation** | | | | | **Patient Allocation** | | | | |
| **Scenario 1** | | | | | | | | | | | |
| TWICEBEE | DL 1 | **0.75** | **0.45** | **0.16** | 0.02 | 0.00 | **0.74** | **0.47** | **0.36** | 0.27 | 0.21 |
| | DL 2 | 0.25 | 0.54 | 0.77 | **0.70** | **0.28** | 0.26 | 0.51 | 0.50 | **0.51** | **0.38** |
| | DL 3 | 0.00 | 0.01 | 0.07 | 0.28 | 0.64 | 0.00 | 0.03 | 0.13 | 0.21 | 0.34 |
| | DL 4 | 0.00 | 0.00 | 0.00 | 0.00 | 0.07 | 0.00 | 0.00 | 0.00 | 0.01 | 0.06 |
| Two-stage | DL 1 | **0.83** | **0.78** | **0.73** | 0.59 | 0.58 | **0.77** | **0.75** | **0.77** | 0.73 | 0.70 |
| | DL 2 | 0.15 | 0.17 | 0.20 | **0.27** | **0.31** | 0.20 | 0.23 | 0.19 | **0.22** | **0.24** |
| | DL 3 | 0.02 | 0.04 | 0.06 | 0.11 | 0.09 | 0.03 | 0.02 | 0.04 | 0.04 | 0.05 |
| | DL 4 | 0.00 | 0.00 | 0.02 | 0.03 | 0.03 | 0.00 | 0.00 | 0.00 | 0.01 | 0.01 |
| DIETE | DL 1 | **0.81** | **0.08** | **0.08** | 0.08 | 0.08 | **0.54** | **0.45** | **0.20** | 0.15 | 0.05 |
| | DL 2 | 0.17 | 0.41 | 0.41 | **0.41** | **0.41** | 0.20 | 0.20 | 0.31 | **0.05** | **0.00** |
| | DL 3 | 0.03 | 0.18 | 0.18 | 0.18 | 0.18 | 0.03 | 0.03 | 0.12 | 0.20 | 0.00 |
| | DL 4 | 0.00 | 0.02 | 0.02 | 0.02 | 0.02 | 0.00 | 0.00 | 0.01 | 0.07 | 0.15 |
| IP-CRM | DL 1 | **0.64** | **0.64** | **0.64** | 0.64 | 0.64 | **0.70** | **0.67** | **0.69** | 0.68 | 0.66 |
| | DL 2 | 0.33 | 0.33 | 0.33 | **0.33** | **0.33** | 0.27 | 0.30 | 0.22 | **0.25** | **0.26** |
| | DL 3 | 0.03 | 0.03 | 0.03 | 0.03 | 0.03 | 0.03 | 0.04 | 0.08 | 0.04 | 0.06 |
| | DL 4 | 0.00 | 0.00 | 0.00 | 0.00 | 0.00 | 0.00 | 0.00 | 0.00 | 0.03 | 0.02 |
| **Scenario 2** | | | | | | | | | | | |
| TWICEBEE | DL 1 | **0.60** | **0.29** | 0.08 | 0.01 | 0.00 | **0.66** | **0.39** | 0.31 | 0.23 | 0.19 |
| | DL 2 | 0.40 | 0.69 | **0.83** | **0.67** | **0.25** | 0.34 | 0.57 | **0.50** | **0.48** | **0.37** |
| | DL 3 | 0.00 | 0.02 | 0.09 | 0.32 | 0.67 | 0.00 | 0.04 | 0.19 | 0.27 | 0.37 |
| | DL 4 | 0.00 | 0.00 | 0.00 | 0.01 | 0.08 | 0.00 | 0.00 | 0.00 | 0.02 | 0.07 |
| Two-stage | DL 1 | **0.72** | **0.67** | 0.52 | 0.56 | 0.54 | **0.68** | **0.63** | 0.62 | 0.57 | 0.56 |
| | DL 2 | 0.22 | 0.25 | **0.34** | **0.31** | **0.34** | 0.26 | 0.32 | **0.29** | **0.34** | **0.34** |
| | DL 3 | 0.05 | 0.06 | 0.10 | 0.10 | 0.08 | 0.06 | 0.04 | 0.09 | 0.07 | 0.09 |
| | DL 4 | 0.01 | 0.02 | 0.04 | 0.03 | 0.04 | 0.00 | 0.00 | 0.00 | 0.02 | 0.01 |
| DIETE | DL 1 | **0.71** | **0.05** | 0.05 | 0.05 | 0.05 | **0.54** | **0.47** | 0.20 | 0.14 | 0.05 |
| | DL 2 | 0.24 | 0.39 | **0.39** | **0.39** | **0.39** | 0.25 | 0.25 | **0.36** | **0.07** | **0.01** |
| | DL 3 | 0.05 | 0.29 | 0.29 | 0.29 | 0.29 | 0.04 | 0.04 | 0.16 | 0.25 | 0.00 |
| | DL 4 | 0.00 | 0.05 | 0.05 | 0.05 | 0.05 | 0.00 | 0.00 | 0.03 | 0.10 | 0.19 |
| IP-CRM | DL 1 | **0.42** | **0.42** | 0.42 | 0.42 | 0.42 | **0.60** | **0.57** | 0.58 | 0.55 | 0.55 |
| | DL 2 | 0.51 | 0.51 | **0.51** | **0.51** | **0.51** | 0.34 | 0.36 | **0.29** | **0.33** | **0.33** |
| | DL 3 | 0.07 | 0.07 | 0.07 | 0.07 | 0.07 | 0.06 | 0.07 | 0.12 | 0.07 | 0.10 |
| | DL 4 | 0.00 | 0.00 | 0.00 | 0.00 | 0.00 | 0.00 | 0.00 | 0.01 | 0.04 | 0.03 |

**Table A.8**: Dose recommendation and patient allocation for Scenarios 1 and 2 for the independent simulation. The bolded entries represent the true MTDC.

| Method | Dose Level | Cycle 1 | Cycle 2 | Cycle 3 | Cycle 4 | Cycle 5 | Cycle 1 | Cycle 2 | Cycle 3 | Cycle 4 | Cycle 5 |
|---|---|---|---|---|---|---|---|---|---|---|---|
| | | **Dose Recommendation** | | | | | **Patient Allocation** | | | | |
| **Scenario 3** | | | | | | | | | | | |
| TWICEBEE | DL 1 | **0.53** | 0.23 | 0.06 | 0.00 | 0.00 | **0.62** | 0.34 | 0.27 | 0.21 | 0.18 |
| | DL 2 | 0.47 | **0.73** | **0.83** | **0.62** | **0.21** | 0.38 | **0.60** | **0.51** | **0.47** | **0.38** |
| | DL 3 | 0.01 | 0.03 | 0.11 | 0.37 | 0.70 | 0.00 | 0.07 | 0.23 | 0.29 | 0.36 |
| | DL 4 | 0.00 | 0.00 | 0.00 | 0.00 | 0.09 | 0.00 | 0.00 | 0.00 | 0.03 | 0.07 |
| Two-stage | DL 1 | **0.60** | 0.51 | 0.53 | 0.51 | 0.52 | **0.59** | 0.51 | 0.51 | 0.48 | 0.48 |
| | DL 2 | 0.30 | **0.36** | **0.34** | **0.37** | **0.36** | 0.30 | **0.40** | **0.35** | **0.38** | **0.38** |
| | DL 3 | 0.09 | 0.08 | 0.09 | 0.09 | 0.10 | 0.09 | 0.08 | 0.13 | 0.12 | 0.12 |
| | DL 4 | 0.00 | 0.04 | 0.04 | 0.03 | 0.02 | 0.01 | 0.01 | 0.01 | 0.02 | 0.02 |
| DIETE | DL 1 | **0.62** | 0.03 | 0.03 | 0.03 | 0.03 | **0.51** | 0.47 | 0.18 | 0.16 | 0.06 |
| | DL 2 | 0.30 | **0.33** | **0.33** | **0.33** | **0.33** | 0.29 | **0.29** | **0.41** | **0.08** | **0.01** |
| | DL 3 | 0.08 | 0.38 | 0.38 | 0.38 | 0.38 | 0.07 | 0.07 | 0.19 | 0.29 | 0.00 |
| | DL 4 | 0.01 | 0.10 | 0.10 | 0.10 | 0.10 | 0.01 | 0.01 | 0.05 | 0.13 | 0.23 |
| IP-CRM | DL 1 | **0.27** | 0.27 | 0.27 | 0.27 | 0.27 | **0.52** | 0.46 | 0.46 | 0.46 | 0.45 |
| | DL 2 | 0.61 | **0.61** | **0.61** | **0.61** | **0.61** | 0.40 | **0.44** | **0.37** | **0.38** | **0.38** |
| | DL 3 | 0.12 | 0.12 | 0.12 | 0.12 | 0.12 | 0.08 | 0.10 | 0.16 | 0.10 | 0.13 |
| | DL 4 | 0.00 | 0.00 | 0.00 | 0.00 | 0.00 | 0.00 | 0.01 | 0.01 | 0.06 | 0.04 |
| **Scenario 4** | | | | | | | | | | | |
| TWICEBEE | DL 1 | **0.54** | **0.23** | 0.05 | 0.00 | 0.00 | **0.69** | **0.39** | 0.32 | 0.21 | 0.16 |
| | DL 2 | 0.46 | 0.72 | **0.77** | **0.52** | 0.11 | 0.31 | 0.56 | **0.49** | **0.48** | 0.36 |
| | DL 3 | 0.01 | 0.05 | 0.18 | 0.46 | **0.70** | 0.00 | 0.05 | 0.19 | 0.28 | **0.38** |
| | DL 4 | 0.00 | 0.00 | 0.00 | 0.01 | 0.19 | 0.00 | 0.00 | 0.00 | 0.03 | 0.11 |
| Two-stage | DL 1 | **0.73** | **0.67** | 0.54 | 0.40 | 0.26 | **0.69** | **0.65** | 0.63 | 0.56 | 0.48 |
| | DL 2 | 0.22 | 0.24 | **0.31** | **0.40** | 0.45 | 0.25 | 0.30 | **0.28** | **0.35** | 0.38 |
| | DL 3 | 0.04 | 0.07 | 0.12 | 0.15 | **0.20** | 0.06 | 0.04 | 0.08 | 0.08 | **0.12** |
| | DL 4 | 0.00 | 0.02 | 0.03 | 0.05 | 0.10 | 0.01 | 0.00 | 0.00 | 0.01 | 0.02 |
| DIETE | DL 1 | **0.75** | **0.04** | 0.04 | 0.04 | 0.04 | **0.54** | **0.47** | 0.20 | 0.15 | 0.04 |
| | DL 2 | 0.20 | 0.33 | **0.33** | **0.33** | 0.33 | 0.24 | 0.24 | **0.37** | **0.06** | 0.01 |
| | DL 3 | 0.05 | 0.31 | 0.31 | 0.31 | **0.31** | 0.04 | 0.04 | 0.15 | 0.25 | **0.00** |
| | DL 4 | 0.00 | 0.08 | 0.08 | 0.08 | 0.08 | 0.00 | 0.00 | 0.03 | 0.10 | 0.21 |
| IP-CRM | DL 1 | **0.23** | **0.23** | 0.23 | 0.23 | 0.23 | **0.50** | **0.48** | 0.48 | 0.45 | 0.42 |
| | DL 2 | 0.60 | 0.60 | **0.60** | **0.60** | 0.60 | 0.40 | 0.41 | **0.35** | **0.39** | 0.40 |
| | DL 3 | 0.17 | 0.17 | 0.17 | 0.17 | **0.17** | 0.10 | 0.11 | 0.16 | 0.11 | **0.14** |
| | DL 4 | 0.01 | 0.01 | 0.01 | 0.01 | 0.01 | 0.00 | 0.01 | 0.01 | 0.05 | 0.04 |

**Table A.9**: Dose recommendation and patient allocation for Scenarios 3 and 4 for the independent simulation. The bolded entries represent the true MTDC.

| Method | Dose Level | Cycle 1 | Cycle 2 | Cycle 3 | Cycle 4 | Cycle 5 | Cycle 1 | Cycle 2 | Cycle 3 | Cycle 4 | Cycle 5 |
|---|---|---|---|---|---|---|---|---|---|---|---|
| | | **Dose Recommendation** | | | | | **Patient Allocation** | | | | |
| | | | | | **Scenario 5** | | | | | | |
| TWICEBEE | DL 1 | **0.31** | 0.10 | 0.02 | 0.00 | 0.00 | **0.61** | 0.34 | 0.25 | 0.16 | 0.10 |
| | DL 2 | 0.67 | **0.79** | **0.67** | 0.34 | 0.06 | 0.39 | **0.58** | **0.50** | 0.42 | 0.30 |
| | DL 3 | 0.02 | 0.10 | 0.31 | **0.63** | **0.70** | 0.00 | 0.08 | 0.25 | **0.37** | **0.45** |
| | DL 4 | 0.00 | 0.00 | 0.00 | 0.03 | 0.24 | 0.00 | 0.00 | 0.01 | 0.04 | 0.15 |
| Two-stage | DL 1 | **0.63** | 0.54 | 0.36 | 0.27 | 0.29 | **0.61** | 0.52 | 0.48 | 0.36 | 0.29 |
| | DL 2 | 0.28 | **0.32** | **0.39** | 0.42 | 0.40 | 0.29 | **0.39** | **0.36** | 0.44 | 0.44 |
| | DL 3 | 0.08 | 0.10 | 0.16 | **0.21** | **0.22** | 0.09 | 0.08 | 0.15 | **0.17** | **0.22** |
| | DL 4 | 0.01 | 0.04 | 0.09 | 0.10 | 0.10 | 0.01 | 0.01 | 0.01 | 0.04 | 0.05 |
| DIETE | DL 1 | **0.63** | 0.02 | 0.02 | 0.02 | 0.02 | **0.52** | 0.47 | 0.18 | 0.13 | 0.03 |
| | DL 2 | 0.28 | **0.23** | **0.23** | 0.23 | 0.23 | 0.29 | **0.29** | **0.40** | 0.07 | 0.01 |
| | DL 3 | 0.07 | 0.39 | 0.39 | **0.39** | **0.39** | 0.06 | 0.06 | 0.19 | **0.30** | **0.00** |
| | DL 4 | 0.01 | 0.19 | 0.19 | 0.19 | 0.19 | 0.00 | 0.00 | 0.04 | 0.15 | 0.28 |
| IP-CRM | DL 1 | **0.07** | 0.07 | 0.07 | 0.07 | 0.07 | **0.38** | 0.34 | 0.33 | 0.30 | 0.26 |
| | DL 2 | 0.54 | **0.54** | **0.54** | 0.54 | 0.54 | 0.42 | **0.45** | **0.39** | 0.41 | 0.42 |
| | DL 3 | 0.35 | 0.35 | 0.35 | **0.35** | **0.35** | 0.18 | 0.19 | 0.25 | **0.21** | **0.24** |
| | DL 4 | 0.03 | 0.03 | 0.03 | 0.03 | 0.03 | 0.01 | 0.02 | 0.04 | 0.08 | 0.08 |
| | | | | | **Scenario 6** | | | | | | |
| TWICEBEE | DL 1 | 0.19 | 0.04 | 0.01 | 0.00 | 0.00 | 0.54 | 0.27 | 0.20 | 0.12 | 0.09 |
| | DL 2 | **0.76** | **0.80** | 0.57 | 0.24 | 0.04 | **0.46** | **0.60** | 0.46 | 0.38 | 0.29 |
| | DL 3 | 0.05 | 0.16 | **0.42** | **0.73** | **0.70** | 0.00 | 0.13 | **0.34** | **0.44** | **0.45** |
| | DL 4 | 0.00 | 0.00 | 0.00 | 0.03 | 0.26 | 0.00 | 0.00 | 0.01 | 0.07 | 0.17 |
| Two-stage | DL 1 | 0.50 | 0.45 | 0.23 | 0.33 | 0.32 | 0.54 | 0.42 | 0.35 | 0.25 | 0.22 |
| | DL 2 | **0.35** | **0.36** | 0.41 | 0.39 | 0.36 | **0.34** | **0.45** | 0.41 | 0.46 | 0.44 |
| | DL 3 | 0.14 | 0.12 | **0.24** | **0.19** | **0.23** | 0.12 | 0.12 | **0.22** | **0.24** | **0.28** |
| | DL 4 | 0.01 | 0.07 | 0.11 | 0.10 | 0.09 | 0.01 | 0.01 | 0.03 | 0.06 | 0.07 |
| DIETE | DL 1 | 0.51 | 0.00 | 0.00 | 0.00 | 0.00 | 0.48 | 0.45 | 0.16 | 0.10 | 0.03 |
| | DL 2 | **0.34** | **0.11** | 0.11 | 0.11 | 0.11 | **0.35** | **0.35** | 0.42 | 0.07 | 0.01 |
| | DL 3 | 0.13 | 0.46 | **0.46** | **0.46** | **0.46** | 0.10 | 0.10 | **0.25** | **0.34** | **0.00** |
| | DL 4 | 0.01 | 0.33 | 0.33 | 0.33 | 0.33 | 0.01 | 0.01 | 0.07 | 0.21 | 0.34 |
| IP-CRM | DL 1 | 0.02 | 0.02 | 0.02 | 0.02 | 0.02 | 0.33 | 0.25 | 0.23 | 0.20 | 0.18 |
| | DL 2 | **0.41** | **0.41** | 0.41 | 0.41 | 0.41 | **0.41** | **0.46** | 0.37 | 0.40 | 0.38 |
| | DL 3 | 0.50 | 0.50 | **0.50** | **0.50** | **0.50** | 0.24 | 0.26 | **0.34** | **0.28** | **0.33** |
| | DL 4 | 0.07 | 0.07 | 0.07 | 0.07 | 0.07 | 0.03 | 0.04 | 0.06 | 0.13 | 0.11 |

**Table A.10**: Dose recommendation and patient allocation for Scenarios 5 and 6 for the independent simulation. The bolded entries represent the true MTDC.

| Method | Dose Level | Cycle 1 | Cycle 2 | Cycle 3 | Cycle 4 | Cycle 5 | Cycle 1 | Cycle 2 | Cycle 3 | Cycle 4 | Cycle 5 |
|---|---|---|---|---|---|---|---|---|---|---|---|
| | | **Dose Recommendation** | | | | | **Patient Allocation** | | | | |
| **Scenario 7** | | | | | | | | | | | |
| TWICEBEE | DL 1 | 0.10 | 0.02 | 0.00 | 0.00 | 0.00 | 0.49 | 0.21 | 0.15 | 0.09 | 0.08 |
| | DL 2 | **0.80** | 0.73 | 0.47 | 0.20 | 0.03 | **0.51** | 0.62 | 0.41 | 0.35 | 0.29 |
| | DL 3 | 0.10 | **0.25** | **0.53** | **0.75** | **0.64** | 0.00 | **0.17** | **0.43** | **0.45** | **0.40** |
| | DL 4 | 0.00 | 0.00 | 0.00 | 0.05 | 0.33 | 0.00 | 0.00 | 0.01 | 0.10 | 0.23 |
| Two-stage | DL 1 | 0.40 | 0.32 | 0.34 | 0.34 | 0.35 | 0.46 | 0.29 | 0.24 | 0.19 | 0.16 |
| | DL 2 | **0.36** | 0.39 | 0.36 | 0.33 | 0.34 | **0.36** | 0.49 | 0.40 | 0.43 | 0.41 |
| | DL 3 | 0.20 | **0.18** | **0.21** | **0.22** | **0.21** | 0.16 | **0.19** | **0.30** | **0.29** | **0.32** |
| | DL 4 | 0.03 | 0.11 | 0.09 | 0.11 | 0.10 | 0.02 | 0.04 | 0.05 | 0.10 | 0.10 |
| DIETE | DL 1 | 0.42 | 0.00 | 0.00 | 0.00 | 0.00 | 0.44 | 0.42 | 0.11 | 0.10 | 0.03 |
| | DL 2 | **0.34** | 0.07 | 0.07 | 0.07 | 0.07 | **0.35** | 0.35 | 0.45 | 0.09 | 0.01 |
| | DL 3 | 0.22 | **0.42** | **0.42** | **0.42** | **0.42** | 0.14 | **0.14** | **0.27** | **0.36** | **0.00** |
| | DL 4 | 0.03 | 0.45 | 0.45 | 0.45 | 0.45 | 0.01 | 0.01 | 0.10 | 0.24 | 0.37 |
| IP-CRM | DL 1 | 0.01 | 0.01 | 0.01 | 0.01 | 0.01 | 0.29 | 0.17 | 0.15 | 0.14 | 0.14 |
| | DL 2 | **0.25** | 0.25 | 0.25 | 0.25 | 0.25 | **0.38** | 0.45 | 0.34 | 0.35 | 0.34 |
| | DL 3 | 0.62 | **0.62** | **0.62** | **0.62** | **0.62** | 0.29 | **0.32** | **0.41** | **0.34** | **0.37** |
| | DL 4 | 0.12 | 0.12 | 0.12 | 0.12 | 0.12 | 0.04 | 0.06 | 0.09 | 0.16 | 0.15 |
| **Scenario 8** | | | | | | | | | | | |
| TWICEBEE | DL 1 | 0.02 | 0.00 | 0.00 | 0.00 | 0.00 | 0.52 | 0.26 | 0.14 | 0.05 | 0.01 |
| | DL 2 | **0.47** | 0.22 | 0.06 | 0.00 | 0.00 | **0.48** | 0.53 | 0.40 | 0.25 | 0.08 |
| | DL 3 | 0.51 | **0.78** | 0.82 | 0.42 | 0.09 | 0.00 | **0.21** | 0.42 | 0.48 | 0.28 |
| | DL 4 | 0.00 | 0.00 | **0.12** | **0.57** | **0.91** | 0.00 | 0.00 | **0.04** | **0.22** | **0.62** |
| Two-stage | DL 1 | 0.54 | 0.30 | 0.20 | 0.39 | 0.62 | 0.57 | 0.38 | 0.26 | 0.12 | 0.06 |
| | DL 2 | **0.32** | 0.39 | 0.23 | 0.10 | 0.04 | **0.32** | 0.47 | 0.40 | 0.36 | 0.21 |
| | DL 3 | 0.12 | **0.19** | 0.31 | 0.21 | 0.10 | 0.11 | **0.14** | 0.29 | 0.35 | 0.38 |
| | DL 4 | 0.02 | 0.13 | **0.26** | **0.30** | **0.24** | 0.01 | 0.02 | **0.05** | **0.17** | **0.35** |
| DIETE | DL 1 | 0.54 | 0.01 | 0.01 | 0.01 | 0.01 | 0.47 | 0.43 | 0.11 | 0.04 | 0.00 |
| | DL 2 | **0.31** | 0.01 | 0.01 | 0.01 | 0.01 | **0.34** | 0.34 | 0.38 | 0.05 | 0.00 |
| | DL 3 | 0.13 | **0.07** | 0.07 | 0.07 | 0.07 | 0.10 | **0.10** | 0.24 | 0.36 | 0.00 |
| | DL 4 | 0.02 | 0.80 | **0.80** | **0.80** | **0.80** | 0.01 | 0.01 | **0.15** | **0.29** | **0.52** |
| IP-CRM | DL 1 | 0.00 | 0.00 | 0.00 | 0.00 | 0.00 | 0.26 | 0.14 | 0.11 | 0.07 | 0.04 |
| | DL 2 | **0.01** | 0.01 | 0.01 | 0.01 | 0.01 | **0.26** | 0.34 | 0.22 | 0.18 | 0.13 |
| | DL 3 | 0.24 | **0.24** | 0.24 | 0.24 | 0.24 | 0.30 | **0.32** | 0.39 | 0.34 | 0.34 |
| | DL 4 | 0.75 | 0.75 | **0.75** | **0.75** | **0.75** | 0.18 | 0.20 | **0.27** | **0.41** | **0.49** |

**Table A.11**: Dose recommendation and patient allocation for Scenarios 7 and 8 for the independent simulation. The bolded entries represent the true MTDC.

| Method | Dose Level | Cycle 1 | Cycle 2 | Cycle 3 | Cycle 4 | Cycle 5 | Cycle 1 | Cycle 2 | Cycle 3 | Cycle 4 | Cycle 5 |
|---|---|---|---|---|---|---|---|---|---|---|---|
| | | **Dose Recommendation** | | | | | **Patient Allocation** | | | | |
| **Scenario 9** | | | | | | | | | | | |
| TWICEBEE | DL 1 | 0.00 | 0.00 | 0.00 | 0.00 | 0.00 | 0.49 | 0.15 | 0.06 | 0.01 | 0.00 |
| | DL 2 | 0.13 | 0.04 | 0.01 | 0.00 | 0.00 | 0.50 | 0.53 | 0.26 | 0.13 | 0.03 |
| | DL 3 | **0.87** | 0.92 | 0.61 | 0.15 | 0.03 | **0.01** | 0.32 | 0.57 | 0.39 | 0.16 |
| | DL 4 | 0.00 | **0.03** | **0.38** | **0.85** | **0.97** | 0.00 | **0.00** | **0.11** | **0.47** | **0.81** |
| Two-stage | DL 1 | 0.33 | 0.31 | 0.39 | 0.73 | 0.88 | 0.47 | 0.17 | 0.10 | 0.03 | 0.01 |
| | DL 2 | 0.34 | 0.23 | 0.08 | 0.02 | 0.01 | 0.33 | 0.53 | 0.27 | 0.17 | 0.07 |
| | DL 3 | **0.25** | 0.24 | 0.22 | 0.06 | 0.02 | **0.17** | 0.24 | 0.48 | 0.34 | 0.26 |
| | DL 4 | 0.08 | **0.23** | **0.31** | **0.19** | **0.09** | 0.02 | **0.07** | **0.16** | **0.46** | **0.66** |
| DIETE | DL 1 | 0.33 | 0.02 | 0.02 | 0.02 | 0.02 | 0.36 | 0.35 | 0.06 | 0.02 | 0.00 |
| | DL 2 | 0.35 | 0.00 | 0.00 | 0.00 | 0.00 | 0.39 | 0.39 | 0.35 | 0.03 | 0.00 |
| | DL 3 | **0.27** | 0.00 | 0.00 | 0.00 | 0.00 | **0.19** | 0.19 | 0.27 | 0.35 | 0.00 |
| | DL 4 | 0.05 | **0.94** | **0.94** | **0.94** | **0.94** | 0.03 | **0.03** | **0.28** | **0.46** | **0.69** |
| IP-CRM | DL 1 | 0.00 | 0.00 | 0.00 | 0.00 | 0.00 | 0.25 | 0.08 | 0.05 | 0.03 | 0.01 |
| | DL 2 | 0.00 | 0.00 | 0.00 | 0.00 | 0.00 | 0.25 | 0.30 | 0.14 | 0.07 | 0.04 |
| | DL 3 | **0.01** | 0.01 | 0.01 | 0.01 | 0.01 | **0.26** | 0.30 | 0.38 | 0.24 | 0.16 |
| | DL 4 | 0.99 | **0.99** | **0.99** | **0.99** | **0.99** | 0.24 | **0.32** | **0.44** | **0.67** | **0.79** |
| **Scenario 10** | | | | | | | | | | | |
| TWICEBEE | DL 1 | 0.00 | 0.00 | 0.00 | 0.00 | 0.00 | 0.52 | 0.07 | 0.03 | 0.01 | 0.00 |
| | DL 2 | 0.03 | 0.01 | 0.00 | 0.00 | 0.00 | 0.46 | 0.54 | 0.13 | 0.07 | 0.02 |
| | DL 3 | 0.94 | 0.88 | 0.44 | 0.07 | 0.01 | 0.02 | 0.38 | 0.65 | 0.24 | 0.13 |
| | DL 4 | **0.03** | **0.10** | **0.56** | **0.93** | **0.99** | **0.00** | **0.00** | **0.19** | **0.69** | **0.85** |
| Two-stage | DL 1 | 0.17 | 0.48 | 0.69 | 0.84 | 0.94 | 0.50 | 0.07 | 0.04 | 0.01 | 0.00 |
| | DL 2 | 0.27 | 0.10 | 0.03 | 0.01 | 0.01 | 0.28 | 0.54 | 0.13 | 0.08 | 0.03 |
| | DL 3 | 0.34 | 0.15 | 0.08 | 0.02 | 0.01 | 0.19 | 0.26 | 0.55 | 0.20 | 0.18 |
| | DL 4 | **0.22** | **0.27** | **0.20** | **0.13** | **0.05** | **0.03** | **0.12** | **0.28** | **0.71** | **0.79** |
| DIETE | DL 1 | 0.33 | 0.03 | 0.03 | 0.03 | 0.03 | 0.30 | 0.30 | 0.03 | 0.01 | 0.00 |
| | DL 2 | 0.23 | 0.00 | 0.00 | 0.00 | 0.00 | 0.35 | 0.35 | 0.31 | 0.02 | 0.00 |
| | DL 3 | 0.32 | 0.00 | 0.00 | 0.00 | 0.00 | 0.26 | 0.26 | 0.30 | 0.31 | 0.00 |
| | DL 4 | **0.13** | **0.92** | **0.92** | **0.92** | **0.92** | **0.07** | **0.07** | **0.35** | **0.59** | **0.79** |
| IP-CRM | DL 1 | 0.00 | 0.00 | 0.00 | 0.00 | 0.00 | 0.25 | 0.04 | 0.02 | 0.01 | 0.01 |
| | DL 2 | 0.00 | 0.00 | 0.00 | 0.00 | 0.00 | 0.25 | 0.29 | 0.06 | 0.03 | 0.01 |
| | DL 3 | 0.00 | 0.00 | 0.00 | 0.00 | 0.00 | 0.25 | 0.30 | 0.35 | 0.12 | 0.12 |
| | DL 4 | **1.00** | **1.00** | **1.00** | **1.00** | **1.00** | **0.25** | **0.38** | **0.57** | **0.84** | **0.86** |

**Table A.12**: Dose recommendation and patient allocation for Scenarios 9 and 10 for the independent simulation. The bolded entries represent the true MTDC.

## A.3 Explain DIETE Design

It is easy to see that, in the DIETE design, the possible dose levels that can be assigned to patients in Cycle 1 are $\{1, \ldots, J\}$. Cycle 2 repeats Cycle 1, assuming for simplicity that patients adapt to the treatment after Cycle 1. Thus, the possible dose levels in Cycle 2 are also $\{1, \ldots, J\}$.

If a DLT is observed for a patient in Cycle 2, that patient may only de-escalate in subsequent cycles until no DLT occurs (at which point the patient exits the trial) or until Cycle 5. Therefore, the possible dose levels for Cycle 5 in this case are $\{1, \ldots, J-3\}$.

On the other hand, if no DLT is observed, the patient may escalate to a higher dose level or to a "second index" dose level. Note that dose level 1 cannot be the second index dose level. Thus, the possible dose levels that a patient (without DLT in Cycle 2) may be assigned in Cycle 3 are $\{2, \ldots, J\}$.

Depending on whether a DLT occurs in Cycle 3, the patient may either de-escalate until no DLT (or until Cycle 5), or escalate until DLT (or until Cycle 5). This leads to possible dose levels of $\{1, \ldots, J-2\}$ for patients with a DLT in Cycle 3, and $\{4, \ldots, J\}$ for those without a DLT in Cycle 3.

Combining these rules, if a patient does not exit the trial before Cycle 5, he or she may be assigned to dose levels $\{1, \ldots, J-2\} \cup \{4, \ldots, J\}$. When $J = 3$, this set does not contain dose level 3.

## A.4 Additional Dependent Simulation Results for Dependent Scenarios

| Method | Dose Level | Cycle 1 | Cycle 2 | Cycle 3 | Cycle 4 | Cycle 5 | Cycle 1 | Cycle 2 | Cycle 3 | Cycle 4 | Cycle 5 |
|---|---|---|---|---|---|---|---|---|---|---|---|
| | | **Dose Recommendation** | | | | | **Patient Allocation** | | | | |
| | | **Scenario 1** | | | | | | | | | |
| TWICEBEE | DL 1 | ***0.81*** | ***0.55*** | ***0.25*** | *0.05* | *0.00* | ***0.75*** | ***0.48*** | ***0.40*** | *0.30* | *0.24* |
| | DL 2 | 0.19 | 0.44 | 0.69 | **0.74** | **0.35** | 0.25 | 0.50 | 0.48 | **0.50** | **0.39** |
| | DL 3 | 0.00 | 0.01 | 0.06 | 0.21 | 0.60 | 0.00 | 0.02 | 0.13 | 0.19 | 0.32 |
| | DL 4 | 0.00 | 0.00 | 0.00 | 0.00 | 0.06 | 0.00 | 0.00 | 0.00 | 0.01 | 0.06 |
| Two-stage | DL 1 | ***0.81*** | ***0.84*** | ***0.77*** | *0.66* | *0.66* | ***0.76*** | ***0.76*** | ***0.78*** | *0.76* | *0.75* |
| | DL 2 | 0.17 | 0.12 | 0.15 | **0.22** | **0.23** | 0.20 | 0.22 | 0.17 | **0.20** | **0.20** |
| | DL 3 | 0.02 | 0.03 | 0.06 | 0.09 | 0.07 | 0.04 | 0.02 | 0.04 | 0.04 | 0.04 |
| | DL 4 | 0.00 | 0.00 | 0.01 | 0.03 | 0.03 | 0.00 | 0.00 | 0.00 | 0.01 | 0.01 |
| DIETE | DL 1 | ***0.81*** | ***0.08*** | ***0.08*** | *0.08* | *0.08* | ***0.55*** | ***0.47*** | ***0.22*** | *0.18* | *0.07* |
| | DL 2 | 0.16 | 0.41 | 0.41 | **0.41** | **0.41** | 0.19 | 0.19 | 0.32 | **0.05** | **0.00** |
| | DL 3 | 0.03 | 0.18 | 0.18 | 0.18 | 0.18 | 0.03 | 0.03 | 0.11 | 0.20 | 0.00 |
| | DL 4 | 0.00 | 0.02 | 0.02 | 0.02 | 0.02 | 0.00 | 0.00 | 0.01 | 0.07 | 0.15 |
| IP-CRM | DL 1 | ***0.73*** | ***0.73*** | ***0.73*** | *0.73* | *0.73* | ***0.77*** | ***0.71*** | ***0.75*** | *0.74* | *0.73* |
| | DL 2 | 0.26 | 0.26 | 0.26 | **0.26** | **0.26** | 0.22 | 0.26 | 0.18 | **0.20** | **0.21** |
| | DL 3 | 0.01 | 0.01 | 0.01 | 0.01 | 0.01 | 0.01 | 0.02 | 0.07 | 0.03 | 0.05 |
| | DL 4 | 0.00 | 0.00 | 0.00 | 0.00 | 0.00 | 0.00 | 0.00 | 0.00 | 0.03 | 0.02 |
| | | **Scenario 2** | | | | | | | | | |
| TWICEBEE | DL 1 | ***0.73*** | ***0.43*** | *0.17* | *0.03* | *0.00* | ***0.69*** | ***0.41*** | *0.35* | *0.25* | *0.23* |
| | DL 2 | 0.27 | 0.56 | **0.77** | **0.70** | **0.33** | 0.31 | 0.55 | **0.48** | **0.50** | **0.41** |
| | DL 3 | 0.00 | 0.01 | 0.07 | 0.27 | 0.61 | 0.00 | 0.04 | 0.17 | 0.22 | 0.30 |
| | DL 4 | 0.00 | 0.00 | 0.00 | 0.00 | 0.06 | 0.00 | 0.00 | 0.00 | 0.02 | 0.06 |
| Two-stage | DL 1 | ***0.73*** | ***0.76*** | *0.61* | *0.62* | *0.61* | ***0.68*** | ***0.66*** | *0.67* | *0.64* | *0.64* |
| | DL 2 | 0.23 | 0.17 | **0.26** | **0.27** | **0.27** | 0.26 | 0.29 | **0.25** | **0.28** | **0.28** |
| | DL 3 | 0.04 | 0.05 | 0.09 | 0.08 | 0.09 | 0.06 | 0.04 | 0.08 | 0.07 | 0.07 |
| | DL 4 | 0.01 | 0.01 | 0.04 | 0.02 | 0.03 | 0.01 | 0.00 | 0.00 | 0.01 | 0.01 |
| DIETE | DL 1 | ***0.74*** | ***0.05*** | *0.05* | *0.05* | *0.05* | ***0.54*** | ***0.47*** | *0.20* | *0.16* | *0.07* |
| | DL 2 | 0.21 | 0.39 | **0.39** | **0.39** | **0.39** | 0.23 | 0.23 | **0.35** | **0.06** | **0.01** |
| | DL 3 | 0.05 | 0.27 | 0.27 | 0.27 | 0.27 | 0.04 | 0.04 | 0.15 | 0.24 | 0.00 |
| | DL 4 | 0.00 | 0.05 | 0.05 | 0.05 | 0.05 | 0.00 | 0.00 | 0.02 | 0.10 | 0.18 |
| IP-CRM | DL 1 | ***0.54*** | ***0.54*** | *0.54* | *0.54* | *0.54* | ***0.66*** | ***0.60*** | *0.62* | *0.61* | *0.60* |
| | DL 2 | 0.42 | 0.42 | **0.42** | **0.42** | **0.42** | 0.30 | 0.34 | **0.26** | **0.29** | **0.29** |
| | DL 3 | 0.04 | 0.04 | 0.04 | 0.04 | 0.04 | 0.04 | 0.05 | 0.11 | 0.06 | 0.08 |
| | DL 4 | 0.00 | 0.00 | 0.00 | 0.00 | 0.00 | 0.00 | 0.00 | 0.01 | 0.04 | 0.03 |

**Table A.13**: Dose recommendation and patient allocation for Scenarios 1 and 2 for the dependent simulation. The bolded entries represent the true MTDC under independent assumption. The *italic* entries represent the true MTDC conditional on a DLT in the prior cycle.

| Method | Dose Level | Cycle 1 | Cycle 2 | Cycle 3 | Cycle 4 | Cycle 5 | Cycle 1 | Cycle 2 | Cycle 3 | Cycle 4 | Cycle 5 |
|---|---|---|---|---|---|---|---|---|---|---|---|
| | | **Dose Recommendation** | | | | | **Patient Allocation** | | | | |
| | | **Scenario 3** | | | | | | | | | |
| TWICEBEE | DL 1 | ***0.61*** | *0.34* | *0.12* | *0.02* | *0.00* | ***0.65*** | *0.36* | *0.31* | *0.23* | *0.21* |
| | DL 2 | 0.39 | **0.65** | **0.80** | **0.68** | **0.28** | 0.35 | **0.58** | **0.49** | **0.47** | **0.39** |
| | DL 3 | 0.00 | 0.02 | 0.07 | 0.30 | 0.66 | 0.00 | 0.06 | 0.20 | 0.27 | 0.33 |
| | DL 4 | 0.00 | 0.00 | 0.00 | 0.00 | 0.06 | 0.00 | 0.00 | 0.00 | 0.03 | 0.06 |
| Two-stage | DL 1 | ***0.64*** | *0.58* | *0.59* | *0.61* | *0.61* | ***0.62*** | *0.55* | *0.56* | *0.55* | *0.56* |
| | DL 2 | 0.27 | **0.30** | **0.27** | **0.27** | **0.29** | 0.29 | **0.37** | **0.31** | **0.33** | **0.32** |
| | DL 3 | 0.09 | 0.09 | 0.10 | 0.09 | 0.07 | 0.09 | 0.08 | 0.12 | 0.10 | 0.10 |
| | DL 4 | 0.00 | 0.03 | 0.03 | 0.03 | 0.03 | 0.01 | 0.01 | 0.01 | 0.02 | 0.02 |
| DIETE | DL 1 | ***0.63*** | *0.02* | *0.02* | *0.02* | *0.02* | ***0.52*** | *0.48* | *0.20* | *0.17* | *0.08* |
| | DL 2 | 0.27 | **0.35** | **0.35** | **0.35** | **0.35** | 0.30 | **0.30** | **0.40** | **0.08** | **0.01** |
| | DL 3 | 0.09 | 0.39 | 0.39 | 0.39 | 0.39 | 0.07 | 0.07 | 0.19 | 0.28 | 0.00 |
| | DL 4 | 0.01 | 0.09 | 0.09 | 0.09 | 0.09 | 0.00 | 0.00 | 0.04 | 0.13 | 0.23 |
| IP-CRM | DL 1 | ***0.40*** | *0.40* | *0.40* | *0.40* | *0.40* | ***0.58*** | *0.51* | *0.53* | *0.52* | *0.53* |
| | DL 2 | 0.52 | **0.52** | **0.52** | **0.52** | **0.52** | 0.35 | **0.40** | **0.31** | **0.34** | **0.33** |
| | DL 3 | 0.08 | 0.08 | 0.08 | 0.08 | 0.08 | 0.07 | 0.08 | 0.14 | 0.09 | 0.11 |
| | DL 4 | 0.00 | 0.00 | 0.00 | 0.00 | 0.00 | 0.00 | 0.01 | 0.01 | 0.05 | 0.03 |
| | | **Scenario 4** | | | | | | | | | |
| TWICEBEE | DL 1 | ***0.61*** | ***0.31*** | *0.09* | 0.01 | 0.00 | ***0.70*** | ***0.41*** | *0.34* | 0.24 | 0.18 |
| | DL 2 | 0.39 | 0.66 | **0.78** | ***0.55*** | *0.14* | 0.30 | 0.54 | **0.47** | ***0.47*** | *0.36* |
| | DL 3 | 0.00 | 0.03 | 0.13 | 0.43 | **0.70** | 0.00 | 0.05 | 0.18 | 0.27 | **0.36** |
| | DL 4 | 0.00 | 0.00 | 0.00 | 0.01 | 0.15 | 0.00 | 0.00 | 0.00 | 0.02 | 0.10 |
| Two-stage | DL 1 | ***0.72*** | ***0.74*** | *0.61* | 0.50 | 0.34 | ***0.69*** | ***0.67*** | *0.67* | 0.61 | 0.54 |
| | DL 2 | 0.23 | 0.20 | **0.26** | ***0.34*** | *0.38* | 0.25 | 0.29 | **0.24** | ***0.30*** | *0.34* |
| | DL 3 | 0.04 | 0.05 | 0.09 | 0.12 | **0.19** | 0.06 | 0.04 | 0.08 | 0.07 | **0.10** |
| | DL 4 | 0.01 | 0.01 | 0.04 | 0.04 | 0.09 | 0.00 | 0.00 | 0.00 | 0.01 | 0.01 |
| DIETE | DL 1 | ***0.73*** | ***0.04*** | *0.04* | 0.04 | 0.04 | ***0.54*** | ***0.47*** | *0.22* | 0.17 | 0.06 |
| | DL 2 | 0.22 | 0.35 | **0.35** | ***0.35*** | *0.35* | 0.24 | 0.24 | **0.36** | ***0.06*** | *0.01* |
| | DL 3 | 0.05 | 0.31 | 0.31 | 0.31 | **0.31** | 0.04 | 0.04 | 0.15 | 0.24 | **0.00** |
| | DL 4 | 0.00 | 0.08 | 0.08 | 0.08 | 0.08 | 0.00 | 0.00 | 0.02 | 0.10 | 0.21 |
| IP-CRM | DL 1 | ***0.36*** | ***0.36*** | *0.36* | 0.36 | 0.36 | ***0.55*** | ***0.52*** | *0.54* | 0.53 | 0.50 |
| | DL 2 | 0.53 | 0.53 | **0.53** | ***0.53*** | *0.53* | 0.37 | 0.39 | **0.31** | ***0.34*** | *0.36* |
| | DL 3 | 0.11 | 0.11 | 0.11 | 0.11 | **0.11** | 0.07 | 0.08 | 0.14 | 0.09 | **0.11** |
| | DL 4 | 0.00 | 0.00 | 0.00 | 0.00 | 0.00 | 0.00 | 0.01 | 0.01 | 0.04 | 0.04 |

**Table A.14**: Dose recommendation and patient allocation for Scenarios 3 and 4 for the dependent simulation. The bolded entries represent the true MTDC under independent assumption. The *italic* entries represent the true MTDC conditional on a DLT in the prior cycle.

| Method | Dose Level | Cycle 1 | Cycle 2 | Cycle 3 | Cycle 4 | Cycle 5 | Cycle 1 | Cycle 2 | Cycle 3 | Cycle 4 | Cycle 5 |
|---|---|---|---|---|---|---|---|---|---|---|---|
| | | **Dose Recommendation** | | | | | **Patient Allocation** | | | | |
| **Scenario 5** | | | | | | | | | | | |
| TWICEBEE | DL 1 | ***0.45*** | *0.16* | 0.05 | 0.00 | 0.00 | ***0.63*** | *0.34* | 0.29 | 0.20 | 0.13 |
| | DL 2 | 0.54 | **0.77** | ***0.72*** | *0.41* | *0.11* | 0.37 | **0.59** | ***0.47*** | *0.44* | *0.33* |
| | DL 3 | 0.02 | 0.06 | 0.23 | **0.57** | **0.67** | 0.00 | 0.07 | 0.23 | **0.32** | **0.41** |
| | DL 4 | 0.00 | 0.00 | 0.00 | 0.01 | 0.22 | 0.00 | 0.00 | 0.01 | 0.04 | 0.13 |
| Two-stage | DL 1 | ***0.63*** | *0.60* | 0.48 | 0.37 | 0.34 | ***0.59*** | *0.54* | 0.53 | 0.44 | 0.37 |
| | DL 2 | 0.28 | **0.29** | ***0.34*** | *0.37* | *0.37* | 0.31 | **0.38** | ***0.33*** | *0.39* | *0.40* |
| | DL 3 | 0.09 | 0.09 | 0.13 | **0.18** | **0.21** | 0.09 | 0.08 | 0.14 | **0.14** | **0.19** |
| | DL 4 | 0.01 | 0.02 | 0.06 | 0.08 | 0.09 | 0.01 | 0.01 | 0.01 | 0.04 | 0.04 |
| DIETE | DL 1 | ***0.65*** | *0.02* | 0.02 | 0.02 | 0.02 | ***0.51*** | *0.46* | 0.19 | 0.14 | 0.05 |
| | DL 2 | 0.27 | **0.24** | ***0.24*** | *0.24* | *0.24* | 0.29 | **0.29** | ***0.38*** | *0.07* | *0.01* |
| | DL 3 | 0.08 | 0.35 | 0.35 | **0.35** | **0.35** | 0.06 | 0.06 | 0.19 | **0.28** | **0.00** |
| | DL 4 | 0.01 | 0.21 | 0.21 | 0.21 | 0.21 | 0.00 | 0.00 | 0.04 | 0.14 | 0.27 |
| IP-CRM | DL 1 | ***0.15*** | *0.15* | 0.15 | 0.15 | 0.15 | ***0.43*** | *0.39* | 0.40 | 0.38 | 0.34 |
| | DL 2 | 0.61 | **0.61** | ***0.61*** | *0.61* | *0.61* | 0.43 | **0.45** | ***0.37*** | *0.41* | *0.42* |
| | DL 3 | 0.22 | 0.22 | 0.22 | **0.22** | **0.22** | 0.13 | 0.15 | 0.20 | **0.15** | **0.18** |
| | DL 4 | 0.02 | 0.02 | 0.02 | 0.02 | 0.02 | 0.01 | 0.01 | 0.02 | 0.07 | 0.06 |
| **Scenario 6** | | | | | | | | | | | |
| TWICEBEE | DL 1 | 0.26 | 0.09 | 0.02 | 0.00 | 0.00 | 0.57 | 0.28 | 0.22 | 0.14 | 0.11 |
| | DL 2 | ***0.70*** | ***0.76*** | *0.64* | *0.33* | *0.07* | ***0.42*** | ***0.61*** | *0.46* | *0.39* | *0.31* |
| | DL 3 | 0.04 | 0.15 | **0.34** | **0.65** | **0.69** | 0.00 | 0.11 | **0.31** | **0.39** | **0.42** |
| | DL 4 | 0.00 | 0.00 | 0.00 | 0.03 | 0.24 | 0.00 | 0.00 | 0.01 | 0.07 | 0.17 |
| Two-stage | DL 1 | 0.53 | 0.46 | 0.36 | 0.36 | 0.41 | 0.55 | 0.43 | 0.38 | 0.29 | 0.26 |
| | DL 2 | ***0.31*** | ***0.36*** | *0.35* | *0.38* | *0.30* | ***0.32*** | ***0.43*** | *0.39* | *0.43* | *0.41* |
| | DL 3 | 0.15 | 0.12 | **0.19** | **0.17** | **0.19** | 0.12 | 0.12 | **0.21** | **0.21** | **0.26** |
| | DL 4 | 0.01 | 0.07 | 0.10 | 0.09 | 0.09 | 0.01 | 0.02 | 0.03 | 0.06 | 0.07 |
| DIETE | DL 1 | 0.53 | 0.01 | 0.01 | 0.01 | 0.01 | 0.46 | 0.43 | 0.16 | 0.12 | 0.04 |
| | DL 2 | ***0.33*** | ***0.11*** | *0.11* | *0.11* | *0.11* | ***0.34*** | ***0.34*** | *0.41* | *0.07* | *0.01* |
| | DL 3 | 0.13 | 0.44 | **0.44** | **0.44** | **0.44** | 0.10 | 0.10 | **0.24** | **0.32** | **0.00** |
| | DL 4 | 0.02 | 0.34 | 0.34 | 0.34 | 0.34 | 0.01 | 0.01 | 0.07 | 0.21 | 0.33 |
| IP-CRM | DL 1 | 0.05 | 0.05 | 0.05 | 0.05 | 0.05 | 0.35 | 0.27 | 0.27 | 0.23 | 0.22 |
| | DL 2 | ***0.51*** | ***0.51*** | *0.51* | *0.51* | *0.51* | ***0.42*** | ***0.47*** | *0.38* | *0.41* | *0.40* |
| | DL 3 | 0.39 | 0.39 | **0.39** | **0.39** | **0.39** | 0.21 | 0.23 | **0.30** | **0.24** | **0.28** |
| | DL 4 | 0.06 | 0.06 | 0.06 | 0.06 | 0.06 | 0.02 | 0.03 | 0.06 | 0.12 | 0.10 |

**Table A.15**: Dose recommendation and patient allocation for Scenarios 5 and 6 for the dependent simulation. The bolded entries represent the true MTDC under independent assumption. The *italic* entries represent the true MTDC conditional on a DLT in the prior cycle.

| Method | Dose Level | Cycle 1 | Cycle 2 | Cycle 3 | Cycle 4 | Cycle 5 | Cycle 1 | Cycle 2 | Cycle 3 | Cycle 4 | Cycle 5 |
|---|---|---|---|---|---|---|---|---|---|---|---|
| | | **Dose Recommendation** | | | | | **Patient Allocation** | | | | |
| **Scenario 7** | | | | | | | | | | | |
| TWICEBEE | DL 1 | 0.18 | 0.05 | 0.01 | 0.00 | 0.00 | 0.49 | 0.22 | 0.18 | 0.13 | 0.11 |
| | DL 2 | ***0.76*** | *0.76* | *0.56* | *0.27* | *0.06* | ***0.50*** | *0.62* | *0.42* | *0.36* | *0.30* |
| | DL 3 | 0.07 | **0.19** | **0.42** | **0.69** | **0.69** | 0.00 | **0.15** | **0.40** | **0.42** | **0.40** |
| | DL 4 | 0.00 | 0.00 | 0.00 | 0.04 | 0.26 | 0.00 | 0.00 | 0.01 | 0.09 | 0.19 |
| Two-stage | DL 1 | 0.38 | 0.42 | 0.39 | 0.39 | 0.42 | 0.46 | 0.30 | 0.28 | 0.22 | 0.20 |
| | DL 2 | ***0.36*** | *0.32* | *0.31* | *0.34* | *0.30* | ***0.35*** | *0.48* | *0.37* | *0.42* | *0.40* |
| | DL 3 | 0.21 | **0.17** | **0.21** | **0.18** | **0.21** | 0.17 | **0.18** | **0.30** | **0.26** | **0.30** |
| | DL 4 | 0.05 | 0.10 | 0.08 | 0.08 | 0.07 | 0.02 | 0.04 | 0.05 | 0.10 | 0.10 |
| DIETE | DL 1 | 0.41 | 0.00 | 0.00 | 0.00 | 0.00 | 0.42 | 0.41 | 0.13 | 0.11 | 0.04 |
| | DL 2 | ***0.38*** | *0.06* | *0.06* | *0.06* | *0.06* | ***0.38*** | *0.38* | *0.42* | *0.09* | *0.02* |
| | DL 3 | 0.19 | **0.43** | **0.43** | **0.43** | **0.43** | 0.13 | **0.13** | **0.28** | **0.34** | **0.00** |
| | DL 4 | 0.02 | 0.44 | 0.44 | 0.44 | 0.44 | 0.01 | 0.01 | 0.10 | 0.25 | 0.36 |
| IP-CRM | DL 1 | 0.02 | 0.02 | 0.02 | 0.02 | 0.02 | 0.31 | 0.19 | 0.19 | 0.18 | 0.17 |
| | DL 2 | ***0.37*** | *0.37* | *0.37* | *0.37* | *0.37* | ***0.39*** | *0.45* | *0.35* | *0.37* | *0.36* |
| | DL 3 | 0.53 | **0.53** | **0.53** | **0.53** | **0.53** | 0.27 | **0.30** | **0.38** | **0.30** | **0.34** |
| | DL 4 | 0.08 | 0.08 | 0.08 | 0.08 | 0.08 | 0.03 | 0.05 | 0.09 | 0.15 | 0.13 |
| **Scenario 8** | | | | | | | | | | | |
| TWICEBEE | DL 1 | 0.05 | 0.01 | 0.00 | 0.00 | 0.00 | 0.54 | 0.26 | 0.16 | 0.07 | 0.03 |
| | DL 2 | ***0.54*** | *0.33* | 0.12 | 0.01 | 0.00 | ***0.46*** | *0.55* | 0.40 | 0.28 | 0.13 |
| | DL 3 | 0.41 | **0.66** | *0.78* | 0.48 | 0.13 | 0.00 | **0.19** | *0.40* | 0.45 | 0.28 |
| | DL 4 | 0.00 | 0.01 | **0.10** | ***0.51*** | ***0.87*** | 0.00 | 0.00 | **0.04** | ***0.20*** | ***0.56*** |
| Two-stage | DL 1 | 0.51 | 0.37 | 0.24 | 0.37 | 0.59 | 0.55 | 0.39 | 0.29 | 0.16 | 0.08 |
| | DL 2 | ***0.35*** | *0.34* | 0.25 | 0.10 | 0.06 | ***0.33*** | *0.46* | 0.38 | 0.35 | 0.24 |
| | DL 3 | 0.12 | **0.17** | *0.30* | 0.20 | 0.09 | 0.11 | **0.13** | *0.28* | 0.32 | 0.36 |
| | DL 4 | 0.02 | 0.12 | **0.20** | ***0.34*** | ***0.26*** | 0.01 | 0.02 | **0.05** | ***0.16*** | ***0.32*** |
| DIETE | DL 1 | 0.51 | 0.01 | 0.01 | 0.01 | 0.01 | 0.47 | 0.44 | 0.13 | 0.06 | 0.02 |
| | DL 2 | ***0.34*** | *0.01* | 0.01 | 0.01 | 0.01 | ***0.34*** | *0.34* | 0.39 | 0.05 | 0.01 |
| | DL 3 | 0.14 | **0.07** | *0.07* | 0.07 | 0.07 | 0.10 | **0.10** | *0.23* | 0.36 | 0.00 |
| | DL 4 | 0.01 | 0.81 | **0.81** | ***0.81*** | ***0.81*** | 0.01 | 0.01 | **0.14** | ***0.28*** | ***0.51*** |
| IP-CRM | DL 1 | 0.00 | 0.00 | 0.00 | 0.00 | 0.00 | 0.27 | 0.16 | 0.15 | 0.10 | 0.07 |
| | DL 2 | ***0.06*** | *0.06* | 0.06 | 0.06 | 0.06 | ***0.29*** | *0.36* | 0.23 | 0.22 | 0.17 |
| | DL 3 | 0.34 | **0.34** | *0.34* | 0.34 | 0.34 | 0.30 | **0.31** | *0.39* | 0.33 | 0.34 |
| | DL 4 | 0.60 | 0.60 | **0.60** | ***0.60*** | ***0.60*** | 0.14 | 0.17 | **0.23** | ***0.35*** | ***0.42*** |

**Table A.16**: Dose recommendation and patient allocation for Scenarios 7 and 8 for the dependent simulation. The bolded entries represent the true MTDC under independent assumption. The *italic* entries represent the true MTDC conditional on a DLT in the prior cycle.

| Method | Dose Level | Cycle 1 | Cycle 2 | Cycle 3 | Cycle 4 | Cycle 5 | Cycle 1 | Cycle 2 | Cycle 3 | Cycle 4 | Cycle 5 |
|---|---|---|---|---|---|---|---|---|---|---|---|
| | | **Dose Recommendation** | | | | | **Patient Allocation** | | | | |
| **Scenario 9** | | | | | | | | | | | |
| TWICEBEE | DL 1 | 0.01 | 0.00 | 0.00 | 0.00 | 0.00 | 0.50 | 0.15 | 0.09 | 0.03 | 0.01 |
| | DL 2 | 0.17 | 0.09 | 0.03 | 0.00 | 0.00 | 0.49 | 0.55 | 0.25 | 0.15 | 0.06 |
| | DL 3 | ***0.82*** | *0.89* | 0.65 | 0.17 | 0.05 | ***0.01*** | *0.30* | 0.57 | 0.35 | 0.16 |
| | DL 4 | 0.00 | **0.02** | ***0.32*** | ***0.82*** | ***0.95*** | 0.00 | **0.00** | ***0.10*** | ***0.47*** | ***0.76*** |
| Two-stage | DL 1 | 0.31 | 0.36 | 0.40 | 0.68 | 0.80 | 0.47 | 0.17 | 0.12 | 0.05 | 0.03 |
| | DL 2 | 0.35 | 0.22 | 0.11 | 0.05 | 0.01 | 0.33 | 0.52 | 0.26 | 0.19 | 0.09 |
| | DL 3 | ***0.26*** | *0.21* | 0.22 | 0.07 | 0.05 | ***0.17*** | *0.24* | 0.47 | 0.32 | 0.26 |
| | DL 4 | 0.08 | **0.21** | ***0.27*** | ***0.20*** | ***0.14*** | 0.02 | **0.07** | ***0.15*** | ***0.45*** | ***0.62*** |
| DIETE | DL 1 | 0.33 | 0.02 | 0.02 | 0.02 | 0.02 | 0.36 | 0.35 | 0.07 | 0.03 | 0.01 |
| | DL 2 | 0.35 | 0.00 | 0.00 | 0.00 | 0.00 | 0.38 | 0.38 | 0.36 | 0.03 | 0.01 |
| | DL 3 | ***0.26*** | *0.01* | 0.01 | 0.01 | 0.01 | ***0.20*** | *0.20* | 0.28 | 0.35 | 0.00 |
| | DL 4 | 0.06 | **0.94** | ***0.94*** | ***0.94*** | ***0.94*** | 0.04 | **0.04** | ***0.27*** | ***0.45*** | ***0.68*** |
| IP-CRM | DL 1 | 0.00 | 0.00 | 0.00 | 0.00 | 0.00 | 0.25 | 0.08 | 0.06 | 0.04 | 0.02 |
| | DL 2 | 0.00 | 0.00 | 0.00 | 0.00 | 0.00 | 0.25 | 0.30 | 0.14 | 0.09 | 0.06 |
| | DL 3 | ***0.04*** | *0.04* | 0.04 | 0.04 | 0.04 | ***0.26*** | *0.31* | 0.37 | 0.23 | 0.18 |
| | DL 4 | 0.95 | **0.95** | ***0.95*** | ***0.95*** | ***0.95*** | 0.23 | **0.31** | ***0.43*** | ***0.64*** | ***0.74*** |
| | | **Dose Recommendation** | | | | | **Patient Allocation** | | | | |
| **Scenario 10** | | | | | | | | | | | |
| TWICEBEE | DL 1 | 0.00 | 0.00 | 0.00 | 0.00 | 0.00 | 0.53 | 0.07 | 0.03 | 0.01 | 0.01 |
| | DL 2 | 0.05 | 0.02 | 0.00 | 0.00 | 0.00 | 0.46 | 0.54 | 0.13 | 0.07 | 0.04 |
| | DL 3 | 0.92 | 0.90 | 0.45 | 0.10 | 0.02 | 0.02 | 0.38 | 0.64 | 0.25 | 0.13 |
| | DL 4 | ***0.03*** | ***0.08*** | ***0.55*** | ***0.90*** | ***0.98*** | ***0.00*** | ***0.00*** | ***0.19*** | ***0.67*** | ***0.82*** |
| Two-stage | DL 1 | 0.20 | 0.52 | 0.67 | 0.82 | 0.89 | 0.49 | 0.08 | 0.05 | 0.02 | 0.01 |
| | DL 2 | 0.27 | 0.08 | 0.04 | 0.02 | 0.01 | 0.28 | 0.54 | 0.13 | 0.09 | 0.04 |
| | DL 3 | 0.33 | 0.11 | 0.08 | 0.03 | 0.03 | 0.19 | 0.26 | 0.54 | 0.20 | 0.19 |
| | DL 4 | ***0.20*** | ***0.29*** | ***0.20*** | ***0.13*** | ***0.07*** | ***0.03*** | ***0.12*** | ***0.28*** | ***0.69*** | ***0.77*** |
| DIETE | DL 1 | 0.31 | 0.03 | 0.03 | 0.03 | 0.03 | 0.29 | 0.29 | 0.03 | 0.02 | 0.01 |
| | DL 2 | 0.20 | 0.00 | 0.00 | 0.00 | 0.00 | 0.35 | 0.35 | 0.31 | 0.02 | 0.01 |
| | DL 3 | 0.35 | 0.00 | 0.00 | 0.00 | 0.00 | 0.28 | 0.28 | 0.30 | 0.30 | 0.00 |
| | DL 4 | ***0.14*** | ***0.93*** | ***0.93*** | ***0.93*** | ***0.93*** | ***0.07*** | ***0.07*** | ***0.35*** | ***0.58*** | ***0.78*** |
| IP-CRM | DL 1 | 0.00 | 0.00 | 0.00 | 0.00 | 0.00 | 0.25 | 0.04 | 0.03 | 0.02 | 0.01 |
| | DL 2 | 0.00 | 0.00 | 0.00 | 0.00 | 0.00 | 0.25 | 0.29 | 0.07 | 0.04 | 0.03 |
| | DL 3 | 0.00 | 0.00 | 0.00 | 0.00 | 0.00 | 0.25 | 0.30 | 0.34 | 0.13 | 0.13 |
| | DL 4 | ***1.00*** | ***1.00*** | ***1.00*** | ***1.00*** | ***1.00*** | ***0.25*** | ***0.37*** | ***0.56*** | ***0.81*** | ***0.83*** |

**Table A.17**: Dose recommendation and patient allocation for Scenarios 9 and 10 for the dependent simulation. The bolded entries represent the true MTDC under independent assumption. The *italic* entries represent the true MTDC conditional on a DLT in the prior cycle.

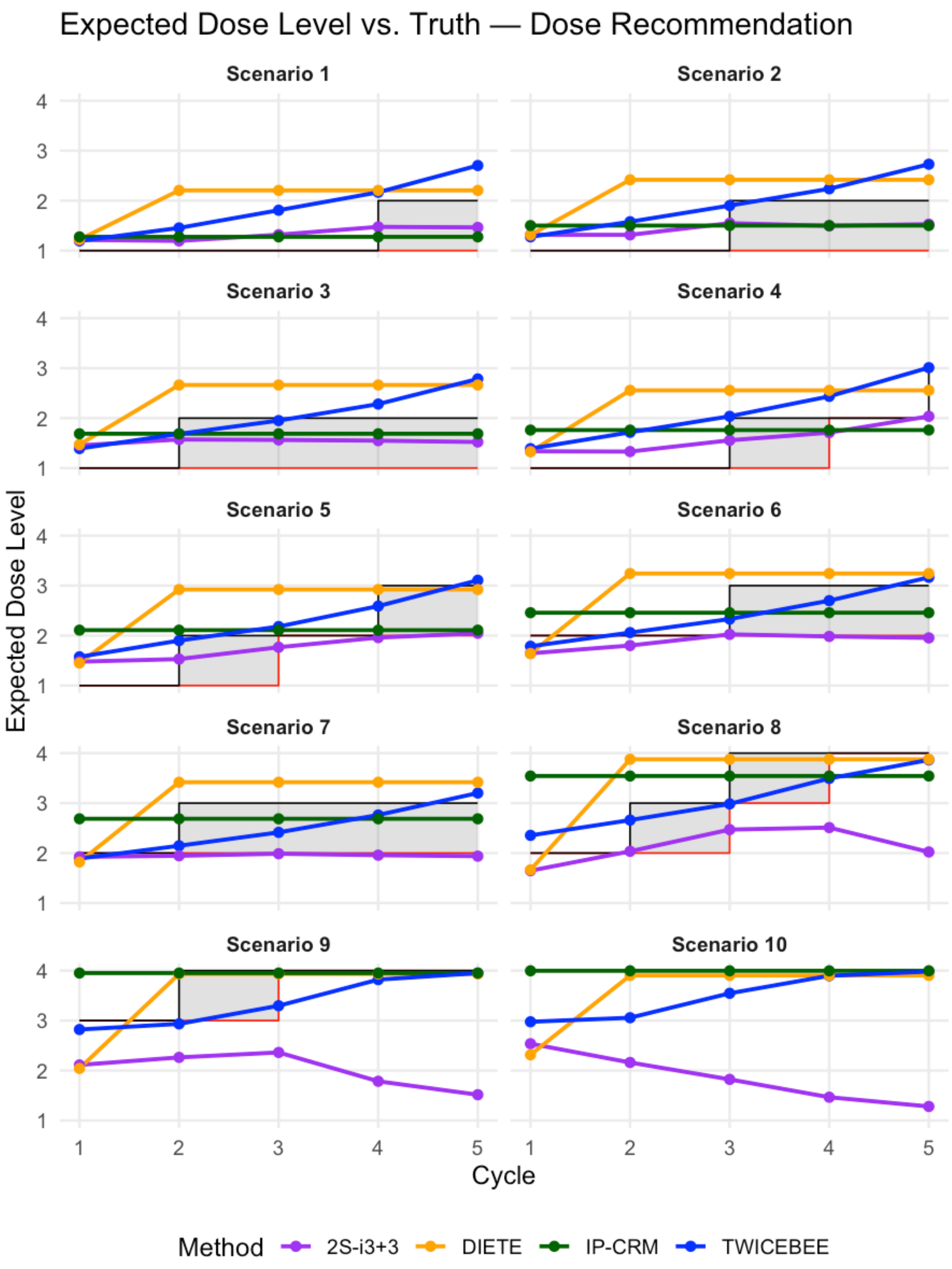

**Figure A.1**: Expected dose levels by cycle for dose recommendations under the dependent simulation. The black line denotes the minimum possible dose levels corresponding to the true MTDS, the red line denotes the maximum possible dose levels corresponding to the true MTDS, and the shaded area between them indicates the "truth region."

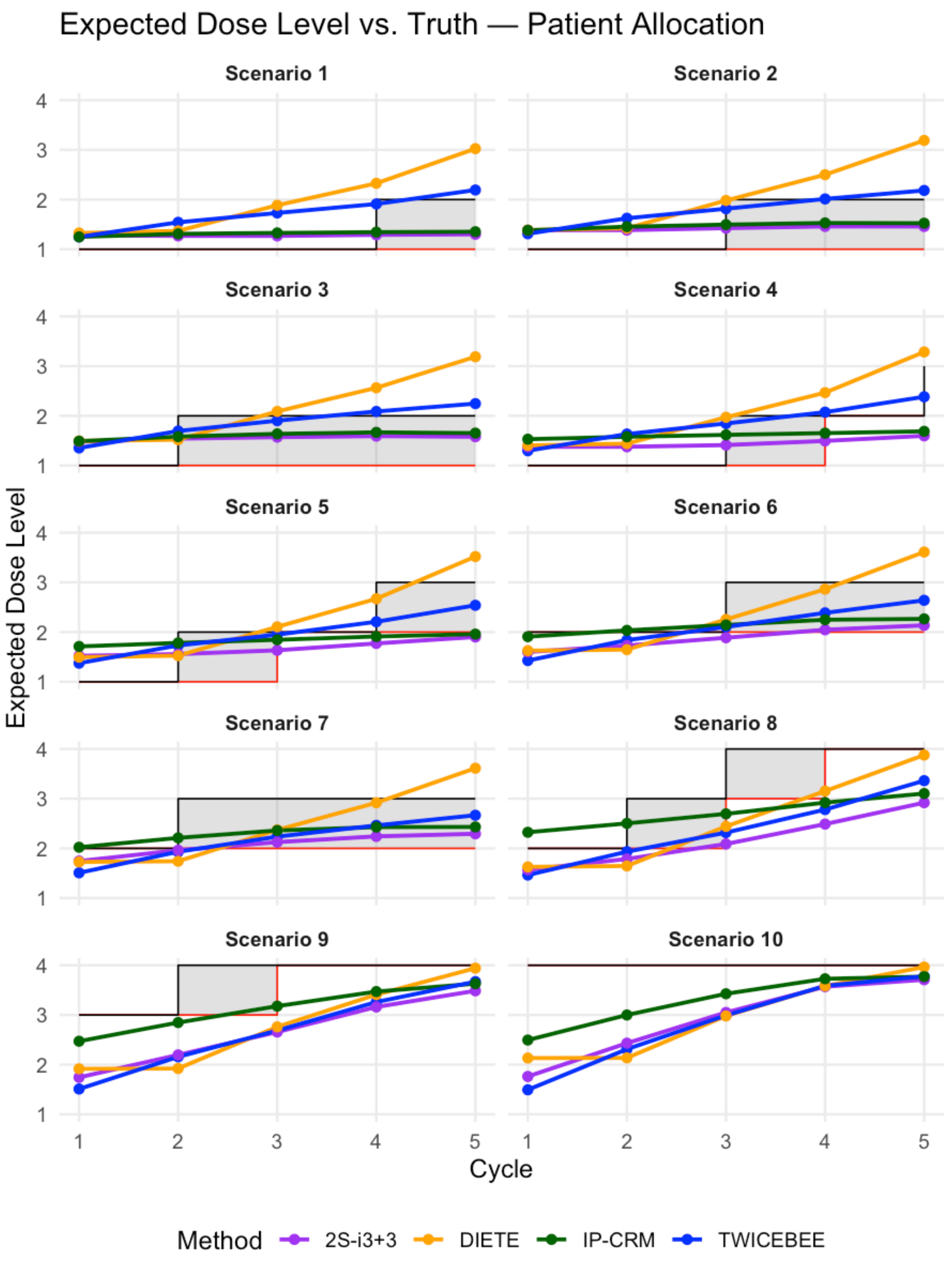

**Figure A.2**: Expected dose levels by cycle for patient allocation under the dependent simulation. The black line denotes the minimum possible dose levels corresponding to the true MTDS, the red line denotes the maximum possible dose levels corresponding to the true MTDS, and the shaded area between them indicates the "truth region."

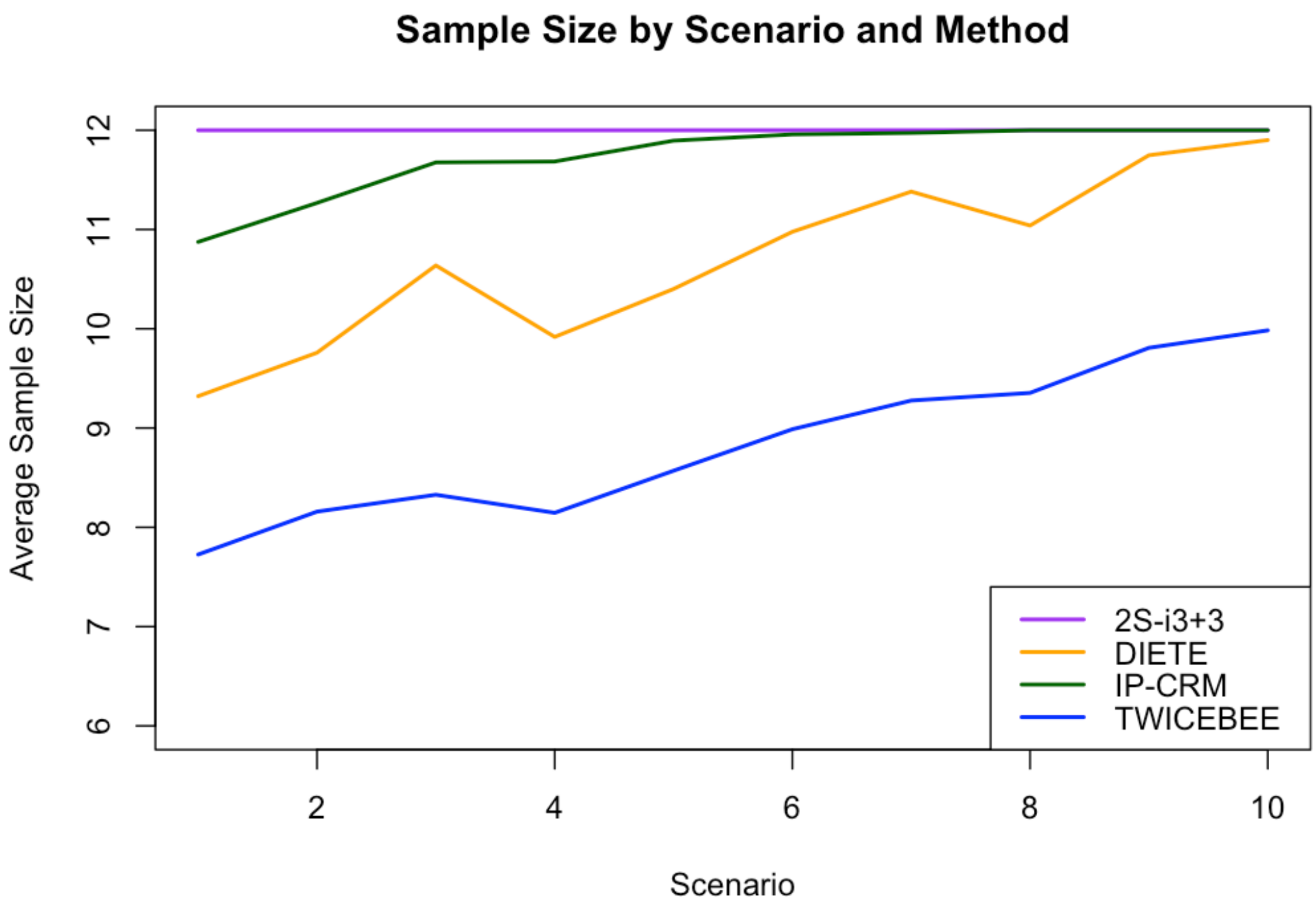

**Figure A.3**: Average sample size of the methods under comparison across the ten simulation scenarios for the dependent scenario.